\documentclass[letterpaper,12pt]{article}
\pdfoutput=1
\usepackage{tabularx} 
\usepackage{amsmath}  
\usepackage{graphicx}
\usepackage[margin=1in,letterpaper]{geometry}
\usepackage{cite}

\usepackage{indentfirst}
\usepackage{txfonts}
\usepackage{subfigure}
\usepackage{xcolor}
\usepackage{url}
\usepackage{multicol}
\usepackage{multirow}
\usepackage{booktabs}
\usepackage{graphicx,subfigure}
\usepackage{hyperref}
\hypersetup{
colorlinks=true,
linkcolor=blue,
anchorcolor=blue,
citecolor=blue,
filecolor=blue,      
urlcolor=black,}

\newcommand{\pk}[1]{\mathbb{P} \left\{ #1 \right\} }

\newcommand{\R}{\mathbb{R}}

\newcommand{\EE}{\mathbb{E}}
\newcommand{\inr}{\in \R}

\newcommand{\COM}[1]{}
\def\N{\mathbb{N}}
\def\IF{\infty}
\def\fracl#1#2{\biggr(\frac{#1}{#2} \biggl) }
\def\E#1{\mathbb{E}\left \{#1 \right\}}
\def\I#1{\mathbb{I}\left (#1 \right)}

\definecolor{codegreen}{rgb}{0,0.6,0}
\definecolor{codegray}{rgb}{0.5,0.5,0.5}
\definecolor{codepurple}{rgb}{0.58,0,0.82}
\definecolor{backcolour}{rgb}{0.95,0.95,0.92}

\definecolor{c20}{rgb}{0.,0.7,0.}
\definecolor{c30}{rgb}{0.,0.,1.}
\definecolor{c40}{rgb}{1,0.1,0.7}
\definecolor{c50}{rgb}{1,0,0}
\definecolor{c60}{rgb}{1,0.9,0.1}
\definecolor{c70}{rgb}{0.50,1.00,0.00}
\def\cl#1{\textcolor{c20}{#1}}
\def\cl#1{#1}

\newtheorem{theo}{Theorem}[section]
\newtheorem{sat}[theo]{Proposition}
\newtheorem{de}[theo]{Definition}
\newtheorem{lem}{Lemma}[section]

\newtheorem{korr}[theo]{Corollary}
\newtheorem{remark}[theo]{Remark}
\numberwithin{equation}{section}

\newcommand{\BQN}{\begin{eqnarray}}
\newcommand{\EQN}{\end{eqnarray}}
\newcommand{\BQNY}{\begin{eqnarray*}}
\newcommand{\EQNY}{\end{eqnarray*}}

\newcommand{\BS}{\begin{sat}}
\newcommand{\ES}{\end{sat}}
\newcommand{\BT}{\begin{theo}}
\newcommand{\ET}{\end{theo}}
\newcommand{\BK}{\begin{korr}}
\newcommand{\EK}{\end{korr}}

\newcommand{\BD}{\begin{de}}
\newcommand{\ED}{\end{de}}
\newcommand{\BIT}{\begin{itemize}}
\newcommand{\EIT}{\end{itemize}}
\newcommand{\BDI}{\begin{description}}
\newcommand{\EDI}{\end{description}}
\newcommand{\BRM}{\begin{remark}}
\newcommand{\ERM}{\end{remark}}
\newcommand{\BEL}{\begin{lem}}
\newcommand{\EEL}{\end{lem}}

\title{Pricing Multi-event Triggered Catastrophe Bonds Based on Copula-POT Model}

\author{YIFAN TANG, CHENGXIU LING , and CONGHUA WEN}

\date{}

\begin{document} 
\maketitle

\begin{abstract}
{The} constantly expanding frequency and loss caused by natural disasters pose a severe challenge to the traditional catastrophe insurance market. {This} paper aims to develop an innovative framework to price catastrophic bonds triggered by multiple events with extreme dependence structure. Given the low contingency of the bond's cash flows and high return of the CAT bond, the multiple-event CAT bond may successfully transfer the catastrophe risk to the big financial markets to meet the diversification of capital allocations for most potential investors. {The} designed hybrid trigger mechanism helps reduce moral hazard and improve bond attractiveness with CIR stochastic rate, displaying the co-movement of the wiped-off coupon, payout principal, the occurrence and intensity of the natural disaster involved.
{As} different triggered indexes of multiple-event catastrophic bonds are heavy-tailed with a variety of dependence relationship, nested Archimedean copulas are introduced with marginal distributions modeled by {POT-GP} distribution for excess data and common parametric models for moderate risks.
{To} illustrate our theoretical pricing framework, we consider a three-event rainstorm CAT bond triggered by catastrophic property losses, in China during 2006--2020. 
{Monte} Carlo simulations are carried out to analyse the sensitivity of the rainstorm CAT bond price in trigger attachment levels, maturity date, catastrophe intensity, and numbers of trigger indicators.

\end{abstract}

\noindent
\textbf{keywords}: extreme value theory; nested Archimedean copula; CAT bond pricing; ARMA model; CIR model

\maketitle

\section{Introduction}
In recent years, the deterioration of the natural environment and growing human activities have increased the level of damage caused by natural disasters, and the economic losses incurred have been on the rise. According to the website \href{https://www.sigma-explorer.com}{Sigma Catastrophe Database}, global catastrophe data during 1970--2021 shows that the frequency of catastrophes has generally been on the rise over the last fifty years. The same goes for losses, with 90$\%$ of all losses over the last decade being in the tens of billions of dollars or more, placing a heavy burden on insurance companies, government finances and society.

Many countries have dispersed catastrophe risk by issuing various types of catastrophe securitization products. For example, in 1997, Hannover Re launched the first successful issue of a CAT bond that included exposure to hurricane and earthquake disasters in Japan, Australia, Canada and Europe. This form filled in a gap in the traditional insurance approach, transferring disaster risks 
to the huge capital market, effectively diversifying catastrophe risk and improving the payment capacity of insurance companies\cite{karagiannis2016modelling}. Among these, catastrophe bonds are considered to be the most mature financial instrument among catastrophe securitization products, attracting an increasing number of investors because of its largely uncorrelation with
the returns of other financial market instruments.

\begin{figure}[htpb]
\centering
\includegraphics[width=15cm]{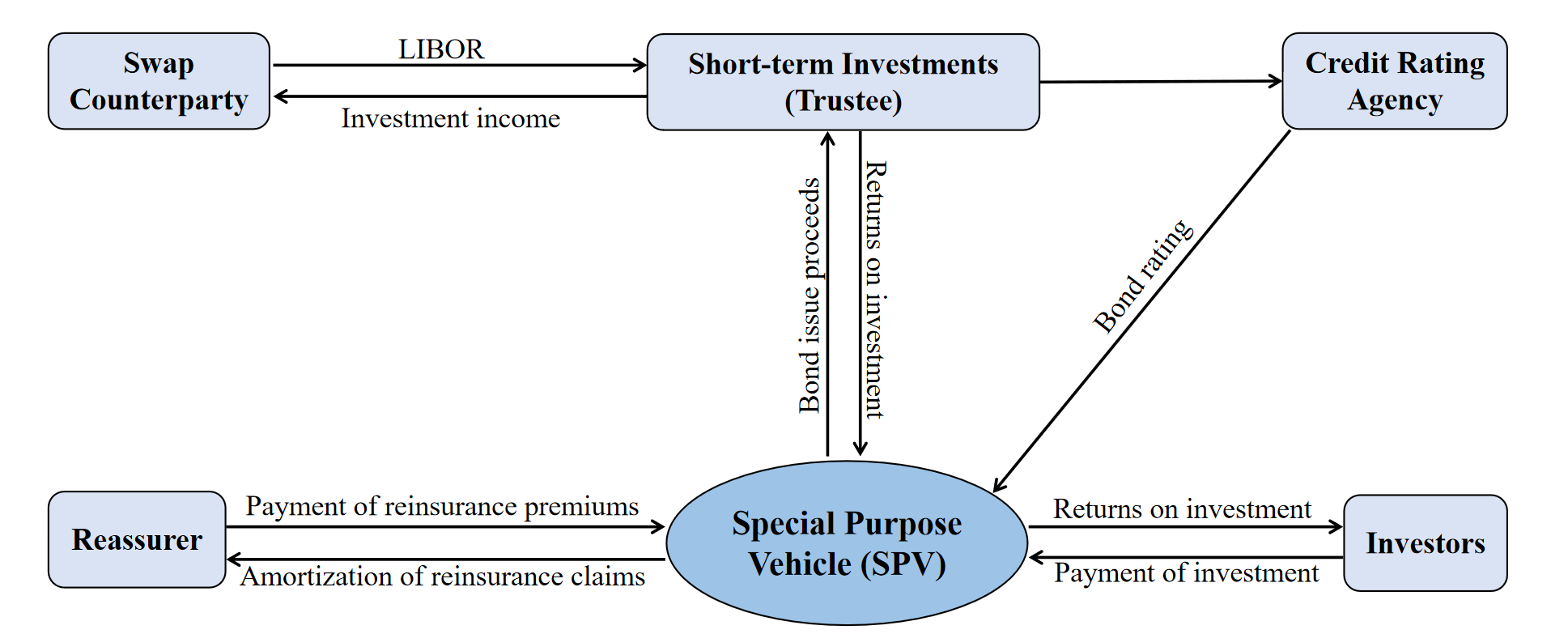}
\caption{Flowchat of Catastrophe bond. 
}\label{Fig:flowchat}
\end{figure}

A CAT bond is a security that pays the
issuer (i.e., collateralized special purpose vehicles (SPVs), ususally established offshore by sponsors who are insurers/reinsurers) when a predefined disaster risk is
realized as showed in Figure \ref{Fig:flowchat}. The SPVs receives (reinsurance) premiums from the investors and reassurers and provides reinsurance coverage in return. The premiums are usually paid to the bond investors as part of coupon payments, which typically also contain a floating portion linked to a certain reference rate, usually the LIBOR, reflecting the return from the trustee. When the specified triggering event occurs, the principal and the coupon payments will be reduced so that some funds can be sent to the sponsors as a reimbursement for the claims paid \cite{cummins2008cat}. An important feature of CAT bonds that tends to differ across the issuer types is the trigger, i.e., the mechanism used to determine when payout must be made to the bond issuer.

As CAT bonds are
designed to cover high layers of insurance losses, a single unprecedented event for instance the accumulated  economic loss caused by catastrophe risk might be appropriate as trigger events\cite{aase1999equilibrium, litzenberger1996assessing,chen2013pricing,nowak2013pricing}.  Although such industry loss trigger mechanisms can completely eliminate basis risk for insurers, the moral hazard is very high, since insurers may exaggerate losses in their loss statistics for their own benefit, creating a significant information asymmetry with capital market investors. Additionally, as argued by \cite{ibrahim2022multiple}, investors' interest in single-event triggered catastrophe bonds is likely to decline in the future, since  the global trend of increasing year-on-year disaster losses and intensity may increase the risk of catastrophe bond claims. Thus, issuing multi-trigger catastrophe bonds could be a solution {using both industry loss indices and physical parameters as triggering conditions\cite{cox2000catastrophe,woo2004catastrophe}. This hybrid trigger mechanism can avoid basis risk and moral hazard with reduced triggering risk, attracting investors with a low risk appetite in the market.} Besides understanding the tail of the marginal trigger indicators  under the framework of extreme value theory\cite{anantapadmanabhan1971some, mcneil1997estimating,zimbidis2007modeling,leppisaari2016modeling,deng2020research}, the dependency among multiple trigger indicators need to be considered through  copula approach\cite{reshetar2008pricing, chao2018multiple, wei2022pricing}. 

This paper aims to develop a CAT bond pricing model with multiple dependent catastrophe risks in a
discrete-time period as an extension of the approach of \cite{cox2000catastrophe}.
The innovation of our trigger mechanism is three-fold. First, the comprehensive selection of catastrophe indicators is examined to cover as many catastrophe indicators as possible to meet the fairness of the bond trigger mechanism setting. Second, the heterogeneity of catastrophic events up to the CAT bond' maturity date is considered under the changing 
global climate. Finally, different from the static wiped-off coupons and principal, our setting of principal-based coupon paid out reflects both the severity and frequency of catastrophe risk.

The remaining of this paper is organized as follows. Section 2 introduces the main methodology establishing the CIR-POT-Copula models based on the independent interest rate risk and natural disaster risk.
Section 3 illustrates our pricing mechanism of main rainstorms in China with nested Archimedean dependence of multiple-event indicators and ARMA-based annual intensity of the rainstorms as well as a series of sensitivity analysis of Monte-carlo simulated bond prices. We conclude this paper in Section 4 with 
extensional discussions on the future research.

\section{Methodology}
\subsection{POT model}
Extreme value theory plays an important role in analysing statistical patterns of extreme value events. There are two typical approaches to extract the extreme samples: the Block Maximum (BM) method and the Peaks-Over-Threshold (POT) model. As the POT model can make full use of the extreme value data in comparison with the BM model, it is widely used in the fields of insurance, hydrology and finance\cite{ma2021return}.

Suppose that $X_1, X_2,\ldots, X_n, \ldots$ is a random sample from parent $X\sim F(x)$, i.e., $X_i$'s are independently and identically distributed with common distribution function (df)  $F(x)$. Given a threshold $u$, the distribution $F_u(y)$ of the excess $Y_{[u]}=X -u$ conditionally on $X>u$, is thus given by
\begin{equation*}
F_u(y)=\pk{X-u \leq y \mid X>u}=\frac{\pk{u<X \leq y+u}}{\pk{X>u}}=\frac{F(y+u)-F(u)}{1-F(u)}.
\end{equation*}
Pickands\cite{pickands31} pointed out that the distribution of the threshold-excess threshold $Y_{[u]}$ can be approximated by generalized Pareto (GP) distribution $G_{\xi, \sigma}(\cdot)$ for sufficiently high threshold, i.e., for the right endpoint $x^*=\sup\{x\inr: F(x)<1\}$
\begin{equation*}
\begin{gathered}
\lim _{u \rightarrow x^{*}} \sup _{0 \leq y \leq x^*-u}\left|F_u(y)-G_{\xi, \sigma}(y)\right|=0. 
\end{gathered}
\end{equation*}
Therefore, the tail distribution function $\overline F(x) = 1-F(x)$ of $X$ can be approximated by
\begin{equation}\label{exceedance}
\overline F(x)=\overline F(u) \overline F_u(x-u) \approx \overline F(u) \overline G_{\xi,\sigma}(x-u), \quad{x}>u.
\end{equation}
Here $\xi\inr$ and $\sigma>0$ are the shape and scale parameters of GP distribution 
$G_{\xi, \sigma}(y)= 1-[1+\xi y / \sigma]^{-1 / \xi}, \  y>0$. In practice, the exceedance probability $\overline F(x)$ gives the insight into the potential risk. Its estimate can be obtained through the extrapolation approach via Eq.\eqref{exceedance}: to get the approximated tail probability of GP model using the maximum likelihood estimation of $\xi,\sigma$ based on excesses $(x_{(i)}-u)'s$ with $x_{(1)}\ge \cdots\ge x_{(n_u)}$ exceeding the threshold $u$ and the estimate of $\overline F(u)$ as $ n_u/n$.  
Theoretically, the threshold $u$ can be determined by minimizing the mean square error of Hill estimator of $\xi$ balancing the model bias and variance. A common graphical approach in the determination of the threshold is to check both the linearity of empirical mean excess function 
\BQN\label{MEF} e_n(u) = \frac{1}{n_u} \sum_{i=1}^{n_u}\left(x_{(i)}-u\right)
\EQN
 and its derived stable estimates of both scale and shape parameters.

\subsection{Copula function theory}
Sklar\cite{sklar1973random} proposed copula as a tool describing the dependence among the marginal variables. Namely, for a joint distribution $G(x_1, \ldots, x_m)$ with marginal df $G_i$ for the $i$th component, the copula $C$ is thus determined by
\begin{equation*}
C(u_1, \ldots, u_m) = G(G_1^{-1}(u_1), \ldots, G_m^{-1}(u_m)),\quad (u_1, \ldots, u_m)\in[0,1]^m. 
\end{equation*}

Table \ref{Table1} lists common Archimedean copula including Clayton, Gumbel, and Frank copula, with a convex and decreasing generator $\phi(t): [0,1]\mapsto (0,\IF)$ satisfying $\phi(1)=0$. Its tail dependence is controlled by the parameter $\theta$ and its copuls is given by 
$$C(u_1, \ldots, u_m) = \phi(\phi^{-1}(u_1), \ldots, \phi^{-1}(u_m)).$$

\begin{table}[htpb]
\begin{center}
\caption{Common Archimedean copulas.}
\label{Table1}
\begin{tabular}{cccc}
\toprule
Copula           & Generator & \multicolumn{1}{c}{$C(u_1, \ldots, u_m)$} & \multicolumn{1}{c}{Parameters} \\ \hline
Gumbel      &                                                                                           $(-\ln t)^{\theta}$ & $\exp \left(-\left(\sum_{i=1}^m\left(-\ln u_i\right)^\theta\right)^{1 / \theta}\right)$                                                               & $\theta \geq 1$                              \\
Clayton     &                                                                                           $\frac{t^{-\theta} -1}\theta$                                                                                              & $\left(\sum_{i=1}^m u_i^{-\theta}-m+1\right)^{-1 / \theta}$                                                               & $\theta \ge 0$                             \\
Frank     &                                                                                           $-\ln\frac{e^{-\theta t}-1}{e^{-\theta}-1}$                                                                                                &             $-\frac{1}{\theta} \ln \left(1+\prod_{i=1}^m \frac{\left(e^{-\theta u_i}-1\right)}{\left(e^{-\theta}-1\right)^{m-1}}\right)$                                                 &                  $\theta \neq 0$                       
 \\ 
\bottomrule
\end{tabular}
\end{center}
\end{table}
Given the analytic tractability of  Archimedean copulas, they are widely applied in  insurance, finance, hydrology and survival analysis etc. In this paper, we consider hierarchical (or nested) Archimedean copulas representing the different dependence among the components \cite{hofert2010sampling}. Namely, the nested Archimedean copula is of form
\BQN\label{nestedArchimedean}
C(u_1, \ldots, u_m) = C_{outer}\left( C_{inner}\left(u_{1}, \ldots, u_k; \theta_1\right), u_{k+1}, \ldots, u_m;\theta_2 \right),
\EQN
where the inner and outer copulas could be one of the three Archimedean copulas in Table \ref{Table1}. The tail dependence could be measured by the  parameters $\theta_1$ and $\theta_2$. As showed in \cite{RePEc:eee:jmvana:v:100:y:2009:i:7:p:1521-1537}, the larger $\theta$ involved in the Gumbel, Clayton or Frank copula indicates a stronger dependence among these variables.

\subsection{Process of counting}
The stochastic process of recording the number of disasters over a certain time period $(0,t]$ is called the counting process $\{N(t), t \geq 0\}$. For all $t\geq 0, N(t)$ is a non-negative integer-valued variable and $N(t)$ is also a non-decreasing function of time $t$. When $0 \leq s < t, N(t) - N(s)$ denotes the number of disasters in the time interval $(s,t]$. A commonly used counting process is the Poisson process, which can be divided into the homogeneous Poisson process and the non-homogeneous Poisson process.

Homogeneous Poisson process has smooth independent increments, when $0\leq t_1< t_2< \cdots < t_k$, the distribution of the increment $N(t_j) - N(t_{j-1})$ depends only on the length of the time interval $\Delta t_j = t_j - t_{j-1}$, not on the specific starting time point of $t_j$, and the increments over non-overlapped intervals are independent of each other. In this case, if we denote by $\lambda$ the average number of disasters in one unit time interval, then
\begin{equation*}
\pk{N(t)=k}=\frac{(\lambda t)^k\exp (-\lambda t)}{k !}, \quad k=0,1,2, \ldots.
\end{equation*}

Non-homogeneous Poisson process (NHPP) allows the instantaneous intensity density $\lambda(t)$  to be a function of $t$. Namely, the  number of disasters up to time $t$  follows Poisson distribution with mean $\Lambda(t)$ satisfying 
\begin{equation*}
\Lambda(t)=\int_0^t \lambda(x) d x.
\end{equation*} 
In general, the number of disasters in the time interval $(s,s+\Delta t]$ follows Poisson distribution with mean 
\begin{equation*}
\Lambda(s+\Delta t)-\Lambda(s)=\int_s^{s+\Delta t} \lambda(x) d x.
\end{equation*}

\subsection{CAT bond pricing model}

This paper considers the pricing of catastrophe bond due to a single disaster, where both the coupon and the principal are at risk in the case of a serious disaster. We refer to the main idea in \cite{chao2018multiple} to consider a coupon paying CAT bond triggered by $m$ dependent catastrophe indicators $x_1, \ldots, x_m$. The investors may receive a portion of the coupon at the end of each year and a portion of the principal back at maturity date. These proportions are determined by the accumulated exceedance of the trigger indicators $x_i$'s over its attachment levels $u_i$'s. Different from \cite{chao2018multiple}, we incorporate a triple of indices $(\alpha_t, \beta_t,\gamma_t)$ into the stressful indicators and the counting process of the disaster $\{N(t),\, t\ge0\}$ (i.e., an integer-valued, nonnegative, and
nondecreasing stochastic process). We will develop a pricing mechanism reflecting the exceedance extent of the trigger indicators and its occurrence of the potential disaster below. We list all symbols involved in Table \ref{Table2}.

\begin{table}[htpb]
\begin{center}
\caption{Notation and its description involved in the pricing formula.}
\label{Table2}
\begin{tabular}{ll}
\toprule
{Notation}     & {Description}            \\ \hline
$F$                              & Principal                                                      \\
$C_t$           & Coupon paid in Year $t,\ t=1, \ldots, T$                                       \\
$R$                              & Coupon rate                                                    \\
$N(t)$                           & Number of disasters up to Year $t$                   \\
$N_t = N(t)-N(t-1)$                    & The number of disasters in Year $t$                             \\
$x_{ij}$                            & The $j$th value of the $i$th trigger indicator up to Year $t$, $j=1,\ldots, N(t)$ \\
$u_i$                     & The attachment level of the $i$th trigger indicator, $i=1,\ldots, m$                     \\
\bottomrule
\end{tabular}
\end{center}
\end{table}
Suppose that the observations of $m$-dimensional indicator vector $(x_{1j}, \ldots, x_{mj}), \, j=N(t-1)+1, \ldots, N(t)$ are independent of the counting process $\{N(t),\, t\ge0\}$, we define the overall catastrophe risk severity  in Year $t$ as below. 
\begin{eqnarray}
\label{Trigger-1}
\alpha_t=f\left(s_{N(t-1)+1},  \ldots, s_{N(t)}\right), \quad s_{j} = \prod_{i=1}^m\left[1-\frac{\left(x_{ij}-u_i\right)_{+}}{x_{ij}}\right],
\end{eqnarray}
where $x_+ = \max(x, 0), f: [0,1]^\N \mapsto [0,1]$, a component-wise non-decreasing, non-negative and right-continuous functional  defined on a filtered physical
probability space.  We
make $f$ a general functional so as to allow for different designs of CAT bonds. In the context of rainstorm CAT bonds, for example, the coupon retention $\alpha_t$ can be
designed to be: 
\begin{itemize}
\item[(i)] The average coupon retention proportion due to the disaster in Year $t$, modeled by 
$$
\alpha_t = \frac{1}{N(t)- N(t-1)}\sum_{j=N(t-1)+1}^{N(t)} \prod_{i=1}^m\left[1-\frac{\left(x_{ij}-u_i\right)_{+}}{x_{ij}}\right].$$
\item[(ii)] The maximum coupon retention proportion due to the disaster in Year $t$, modeled by
$$\alpha_t = \max_{ N(t-1)+1\le j \le N(t)} \prod_{i=1}^m\left[1-\frac{\left(x_{ij}-u_i\right)_{+}}{x_{ij}}\right].$$
\end{itemize}
\BRM \label{Remark1} Note that each $s_j\in(0,1]$. The case with $s_j=1$ tells that in the $j$th disaster, all trigger indicators are below the attachment levels. The smaller the $s_j$ is, the indicator $x_j =(x_{1j}, \ldots, x_{mj})$ is more likely to be far larger than its attachment level $u=(u_1, \ldots, u_m)$.  It follows by the component-wise non-decreasing property of the functional $f$ that the $\alpha_t\in(0,1]$ is appropriate to quantify the coupon retention proportion. 
\ERM

To further trigger partial principal due to more than one indicator's exceedance over its attachment level, we introduce the following two indices  $\beta_t$ and $\gamma_t$ below. Similar to $s_j$'s and the functional $f$ in Eq.\eqref{Trigger-1}, we define 
\begin{eqnarray}
\label{Trigger-2}
\beta_t &=& g\left(s^*_{N(t-1)+1},  \ldots, s^*_{N(t)}\right), \quad s^*_{j} = \prod_{1\le i_1 < i_2\le m}\left[1-\frac{\left(x_{i_1 j}-u_{i_1}\right)_{+}}{x_{i_1j}} \cdot \frac{\left(x_{i_2j}-u_{i_2}\right)_{+}}{x_{i_2 j}}\right], \\
\label{Trigger-3}
\gamma_t &=&h\left(s^{**}_{N(t-1)+1},  \ldots, s^{**}_{N(t)}\right), \quad s^{**}_{j} = \prod_{1\le i_1<i_2<i_3\le m}\left[1-\frac{\left(x_{i_1 j}-u_{i_1}\right)_{+}}{x_{i_1j}}\cdot \frac{\left(x_{i_2 j}-u_{i_2}\right)_{+}}{x_{i_2j}}\cdot \frac{\left(x_{i_3 j}-u_{i_3}\right)_{+}}{x_{i_3 j}}\right],\quad
\end{eqnarray}
where $g$ and $h$ are two functionals similar to $f$, i.e., they are component-wise nondecreasing, non-negative and right-continuous functionals from $[0,1]^\N$ to $[0,1]$. In the following proposition, we introduce the properties of $(s_j, s_j^*, s_j^{**})$'s and $(\alpha_t, \beta_t,\gamma_t)$'s.

\BS
\label{Prop1}
Let $(s_j, s_j^*, s_j^{**})$ be defined by Eq.\eqref{Trigger-1}-\eqref{Trigger-3} and $(\alpha_t, \beta_t, \gamma_t)$'s are three component-wise non-decreasing, non-negative and right-continuous functionals from $[0,1]^\N$ to $[0,1]$. 
\begin{itemize}
\item[(i)] It follows that $s_j, s_j^*, s_j^{**}$ range over $[0,1]$. The case with $s_j^*<1$ ( $s_j^{**}<1$) implies that in the $j$th disaster, among all $m$ trigger indicators, there are at least two (three) indicators above its attachment levels simultaneously. Similar to $s_j$, the values of $s_j^*$ and $s_j^{**}$ describe the exceedances over its thresholds. 
\item[(ii)] Quantitatively, we have $0<s_j \le s_j^* \le s_j^{**}\le1$, which indicates further that 
$$0<\alpha_t \le \beta_t\le \gamma_t\le 1$$
provided that three functionals are taken as the same ones. 
\item[(iii)] It follows by the component-wise non-decreasing monotonicity of the functionals $f,g$ and $h$ that the quantities $\alpha_t, \beta_t, \gamma_t$ reflect not only the frequency but also the severity of the disaster, with lower values indicating a higher frequency and extreme severity of the trigger indicators. 
\end{itemize}
\ES

Based on the properties showed in Proposition \ref{Prop1}, we will design a CAT bond with principal \cl{remained} 
protected unless there are at least two trigger indicators triggered simultaneously, namely $\beta_t<1$. In this case, we differ the wiped-off principal  with proportion of $\beta_t$ and $\beta_t/2$ according to the accumulated proportion  with exactly two and at least three indicators triggered simultaneously.  This protects the investor's principal will never be paid out to zero, thus increasing the investor's interest in the investment of CAT bond. Let the investor buys a CAT bond with a face value of $F$ at Year $t$ and maturity at Year $T$. Denote by $C_{t,s}$ the coupon paid in Year $s$  and $F_{t,T}$ the redemption value at Year $t$. We define
\begin{equation}
\label{Cashflow}
\left\{\begin{array}{lr}
{C_{t,s}}=\left\{\begin{array}{ll}
\alpha_{t+1} \cdot C_0, &s=t+1, \\
\alpha_s \cdot \prod_{k=t+1}^{s-1}\left(\beta_k \cdot \fracl12^{\I{\gamma_{k}<1}}\right) \cdot C_0, & s=t+2, \ldots,T,\\
\end{array}\right.\\
{F_{t,T}}=\prod_{k=t+1}^{T}\left(\beta_k \cdot \fracl12^{\I{\gamma_{k}<1}}\right) \cdot F,
\end{array}\right.
\end{equation}
where $(\alpha_t, \beta_t, \gamma_t)$'s are defined by Eq.\eqref{Trigger-1}-\eqref{Trigger-3} and {$C_0 = F\cdot R$}, the coupon paid according to a fixed coupon rate $R$ for a face value of $F$.

\BRM\label{Remark2}
Our trigger mechanism in Eq.\eqref{Cashflow} with the hierarchical proportional coupon and principals paid out may attract more investors in comparison with the hybrid trigger mechanism given by \cite{wei2022pricing}. Since therein the current and future coupons will be paid out once one of the indicators is triggered and the principal at maturity will be completely wiped out  once both indicators are triggered simultaneously.   
\ERM

Finally, the price of the CAT bond, as the present value of all future cashflows, denoted by $P_t$, is given by
\begin{equation}
\label{PricingMechanism}
P_t=\sum_{s=t+1}^T  \E{C_{t,s}} p(t, s)+ \E{F_{t,T}} \cdot p(t, T),
\end{equation}
where $p(t, s)$ represents the discount factor at time $t$ of the zero-coupon bond with redemption value of 1 at time $s, t<s\le T$, given by
\begin{equation*}
p(t, s)= \EE\left[\exp \left(-\int_t^s r(u) \mathrm{d} u\right)\right].
\end{equation*}

In our context, we consider the Cox-Ingersoll-Ross (CIR) model\cite{cox1985intertemporal} of the continuous-time interest rate process $\{r(t),\, t\ge0\}$. The CIR interest rate model has specific advantages over the Vasicek interest rate\cite{RN30} model is that it is simple, easy to handle, mean-reverting and removes the possibility of rates less than zero. It satisfies the following first-order differential equation under the risk-neutral probability measure:
\begin{equation}\label{CIR}
d r(t)=k\left(\theta-r(t)\right) d t+\varepsilon \sqrt{r(t)} d W_t,
\end{equation}
where $k>0$ is the mean reversion measure, $\varepsilon>0 $ is the volatility parameter, $\theta>0$ is the long-run mean of the interest rate, and \{$W_t, \, t \geq 0\}$ denotes the standard Wiener process. Consequently, the discount factor $p(t, T)$ is given by
\begin{equation*}
p(t, T)=A(t, T) e^{-B(t, T) r(t)},
\end{equation*}
where
\begin{equation*}
\begin{aligned}
A(t, T) &=\left[\frac{2 \eta e^{(\kappa+\eta) (T-t) / 2}}{(\kappa+\eta)\left(e^{\eta (T-t)}-1\right)+2 \eta}\right]^{2 \kappa \theta / \varepsilon^2}, \\
B(t, T) &=\frac{2\left(e^{\eta (T-t)}-1\right)}{(\kappa+\eta)\left(e^{\eta (T-t)}-1\right)+2 \eta}, \\
\eta &=\sqrt{\kappa^2+2 \varepsilon^2}.
\end{aligned}
\end{equation*}

\section{Empirical Analysis}
To illustrate our pricing mechanism, this section focuses on the pricing of severe rainstorm in China based on all recorded 245 main rainstorms in China during 2006--2020 and three main hazard indicators from the website \href{https://data.cnki.net/yearBook/single?id=N2021120059}{\textit{China Statistical Yearbook of Natural Disasters}}. These indicators cover all the disaster indicators, namely, the affected population (AP) accumulating the number of deaths, missing, emergency re-locations and of people with drinking water damaged, crop affected area (CAA) reflecting both cropland flooding area and its harvest-affected area, and  direct economic loss (DEL) adjusted with the Consumer Price Index (CPI) in 2020, summing the damaged losses of collapsed/damaged houses and other properties multiplied by its damaged ratio.

In what follows, we will present first the basic patterns of our triple of trigger indicators $(X_1, X_2, X_3)$ representing AP, CAA and DEL in Section 
\ref{Section-3.1}. The non-normality of these indicators motives us to split the full range of data into bulk  and tail parts using classical parametric models and Peaks-Over-Threshold (POT) 
 approach in Section \ref{Section-3.2}, followed by the investigation of the joint distribution functions for disaster indicators data using an nested Archimedean copula in Section \ref{Section-3.3}. The ARMA model is used to predict disaster intensity over the next 3 years in Section \ref{Section-3.4}. Monte Carlo simulation of the CAT bond price is carried out in Section \ref{Section-3.5}. Here we take the retention functions of $f,h$ and $g$ in Eq.\eqref{Trigger-1}--Eq.\eqref{Trigger-3} as average function unless stated otherwise.

\subsection{Descriptive analysis of trigger indicators}
\label{Section-3.1}

\begin{figure}[htbp!]
  \centering
  \subfigure[]{\includegraphics[width=5cm,height=4cm]{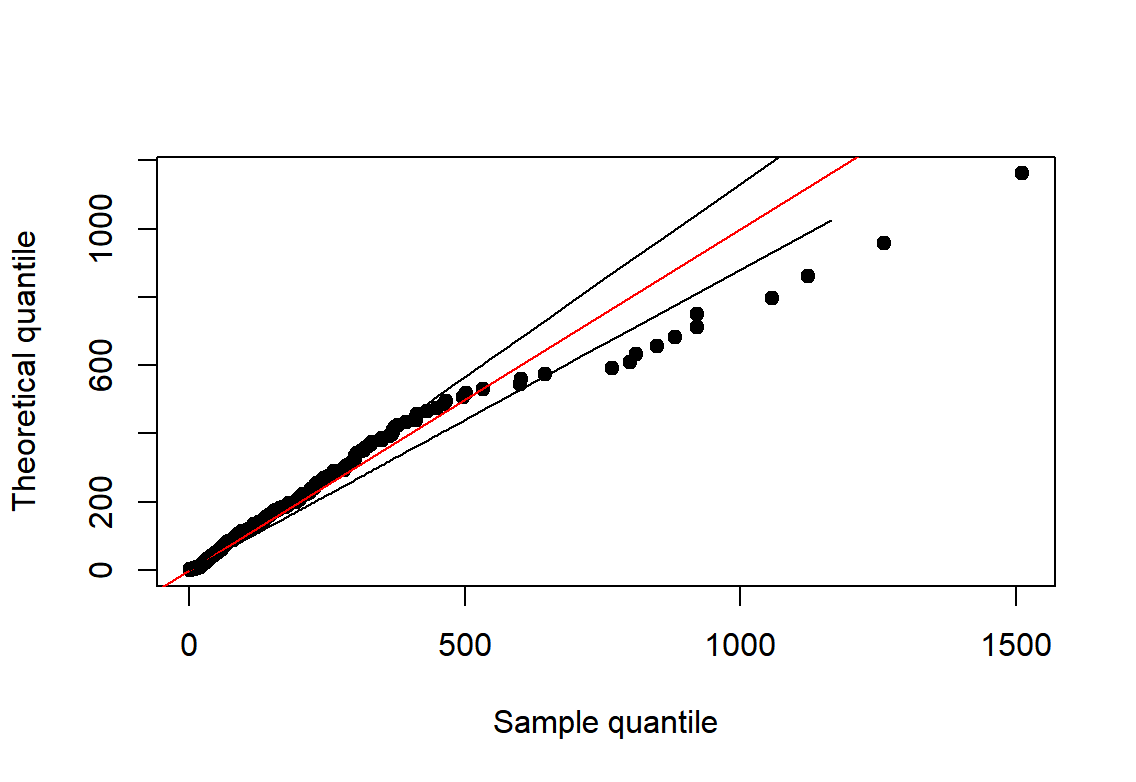}
 }
  \subfigure[]{\includegraphics[width=5cm,height=4cm]{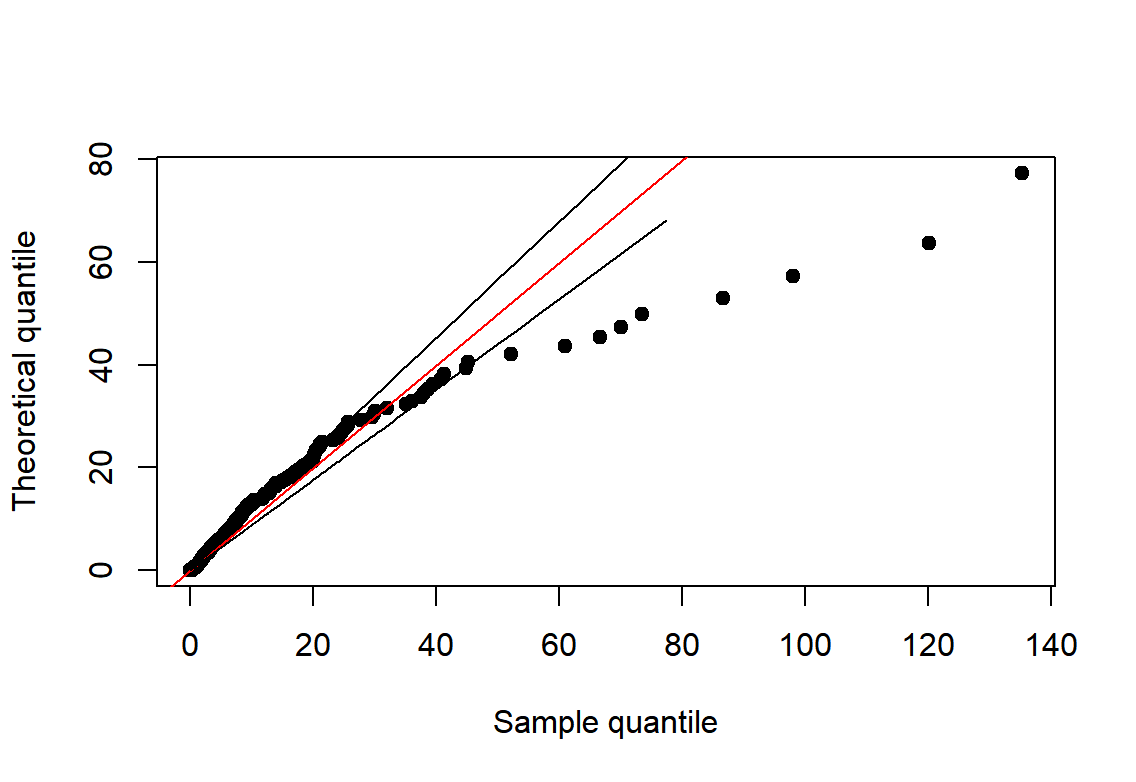}
  \label{Fig3-2}}
    \subfigure[]{\includegraphics[width=5cm,height=4cm]{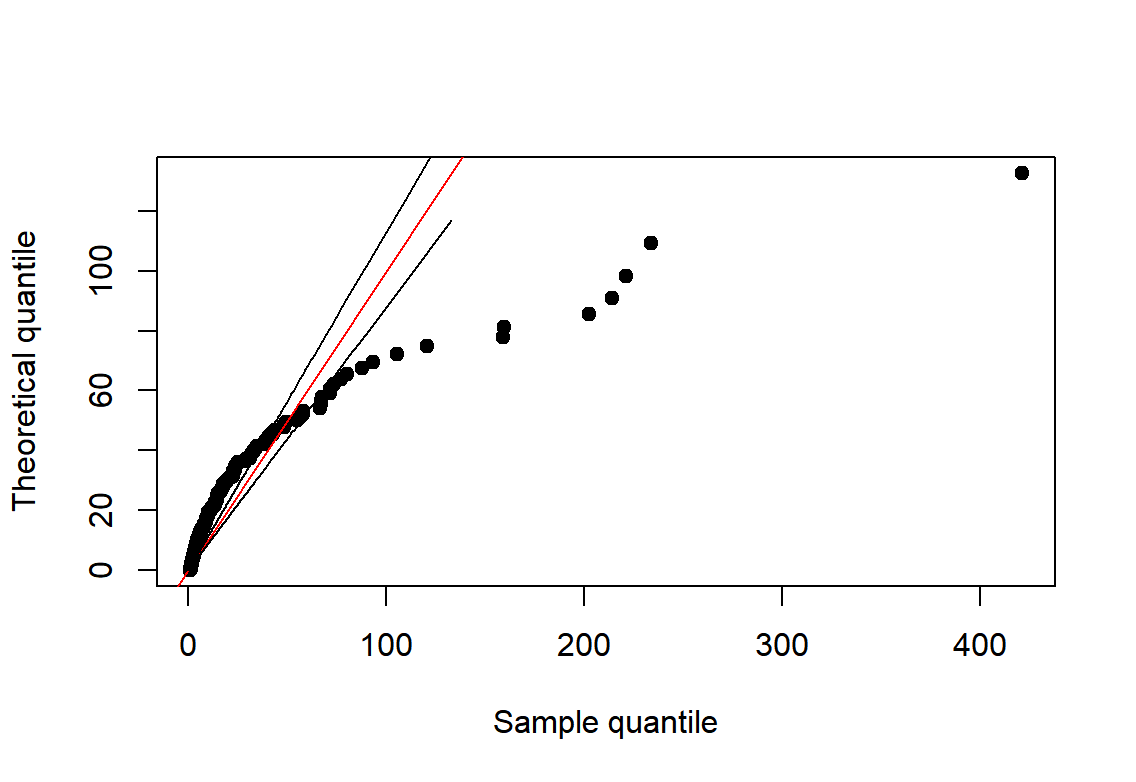}
 }
  \caption{Exponential QQ plots of (a) affected population (AP), (b) crop affected area (CAA)  and (c) directed economic loss (DEL), with the red dashed line representing the exponential reference line.}
  \label{Fig2}
\end{figure}

We see from Table \ref{Table3} that, the averages of all the three main indicators are far larger than its median accordingly. Moreover, both skewness and kurtosis of these indicators, which is far larger than 0 and 3,  indicate the right-skewed pattern of these indicators with possible heavy tails. This is graphically confirmed by the exponential QQ plots in Figure \ref{Fig2} with a downward convex deviation from the theoretical straight line.

\begin{table}[htbp!]
\begin{center}
\caption{Descriptive statistics of affected population (AP) in million, crop affected area (CAA) in million hectares, direct economic loss (DEL) in billion yuan.}
\label{Table3}
\begin{tabular}{ccccccc}
\toprule
Trigger Indicators & Maximum  & Minimum & Mean    & Median  & {Skewness} & {Kurtosis} \\ \hline
AP       & 1510.000 & 1.600   & 187.880 & 116.750 & \textbf{2.771}    & \textbf{12.678}   \\
CAA      & 135.200  & 0.005   & 12.492  & 7.000   & \textbf{3.689}    & \textbf{20.458}   \\
DEL      & 420.816  & 1.011   & 21.411  & 8.083   & \textbf{5.154}    & \textbf{37.823}   \\ 
\bottomrule
\end{tabular}
\end{center}
\end{table}

\subsection{POT-based tail analysis of trigger indicators}
\label{Section-3.2}
\begin{figure}[ht]
  \centering
  \subfigure[]{\includegraphics[width=5.2cm,height=4cm]{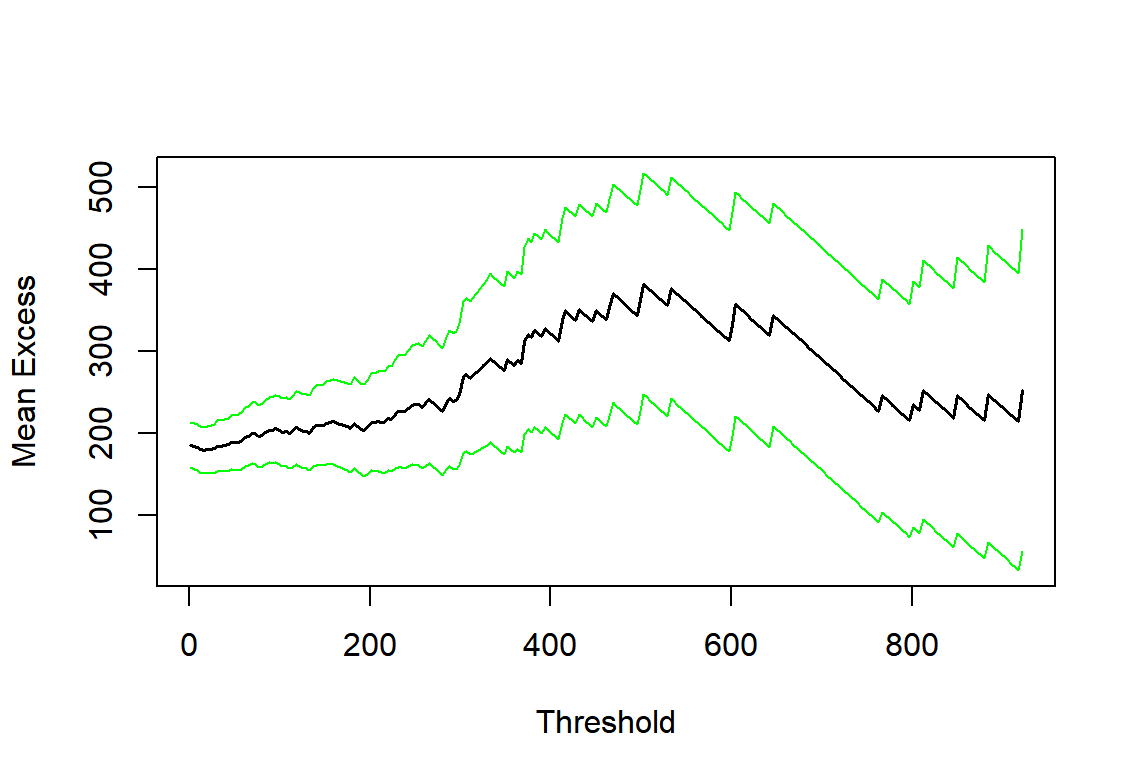}
  }
  \subfigure[]{\includegraphics[width=5.2cm,height=4cm]{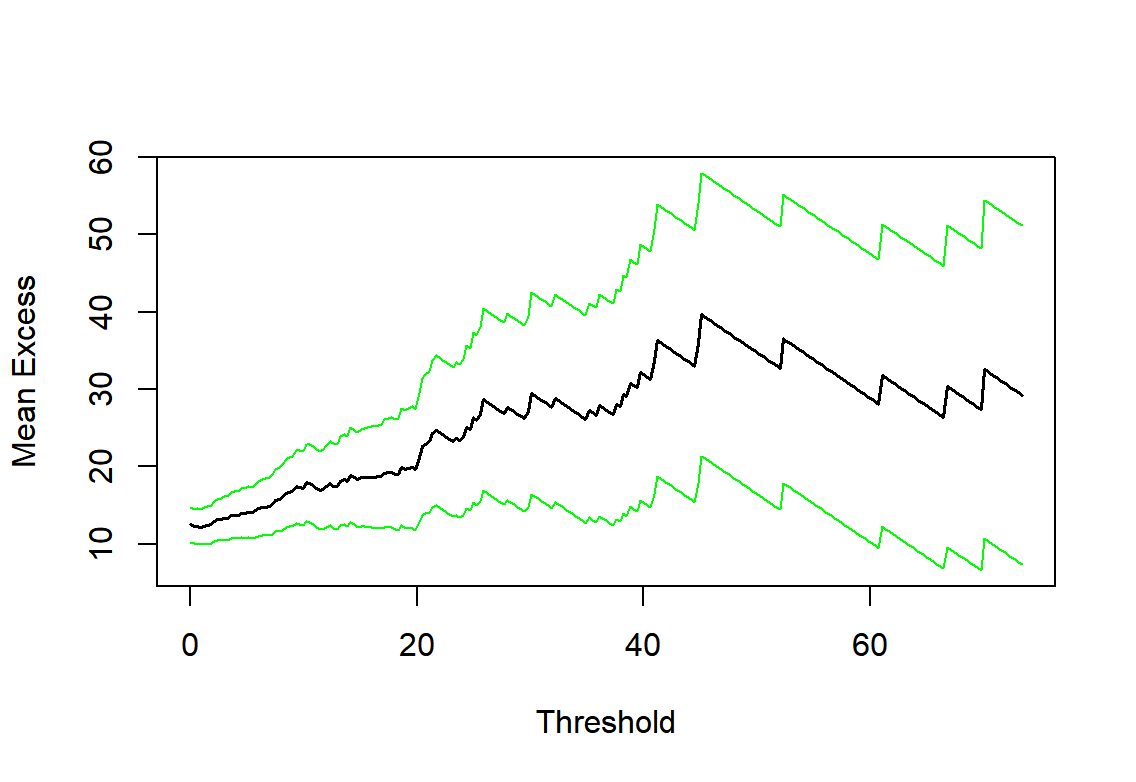}
 }
    \subfigure[]{\includegraphics[width=5.2cm,height=4cm]{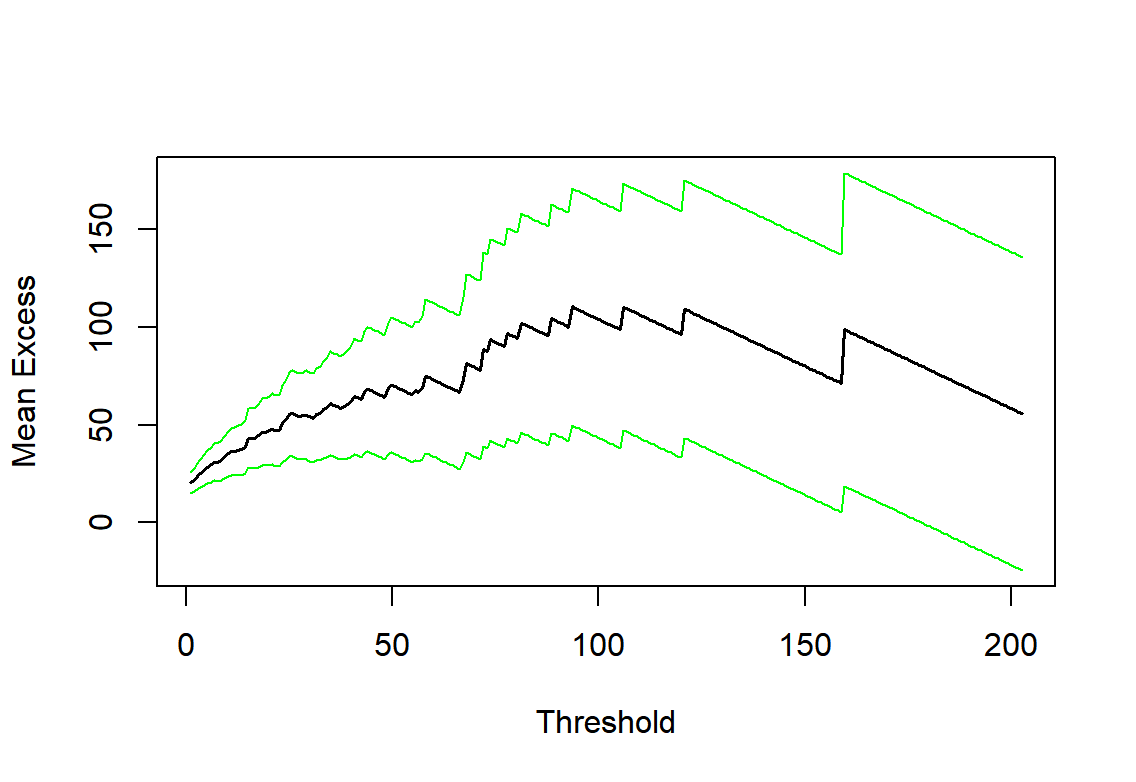}
  }
  \caption{The mean residual life plots for (a) affected population (AP), (b)  crop affected area (CAA) and (c) directed economic loss (DEL) subsequently. The green lines correspond to the 95\% confidence interval.}
  \label{Fig3}
\end{figure}

\begin{figure}[htpb]
  \centering
  \subfigure[]{\includegraphics[width=7cm,height=3.95cm]{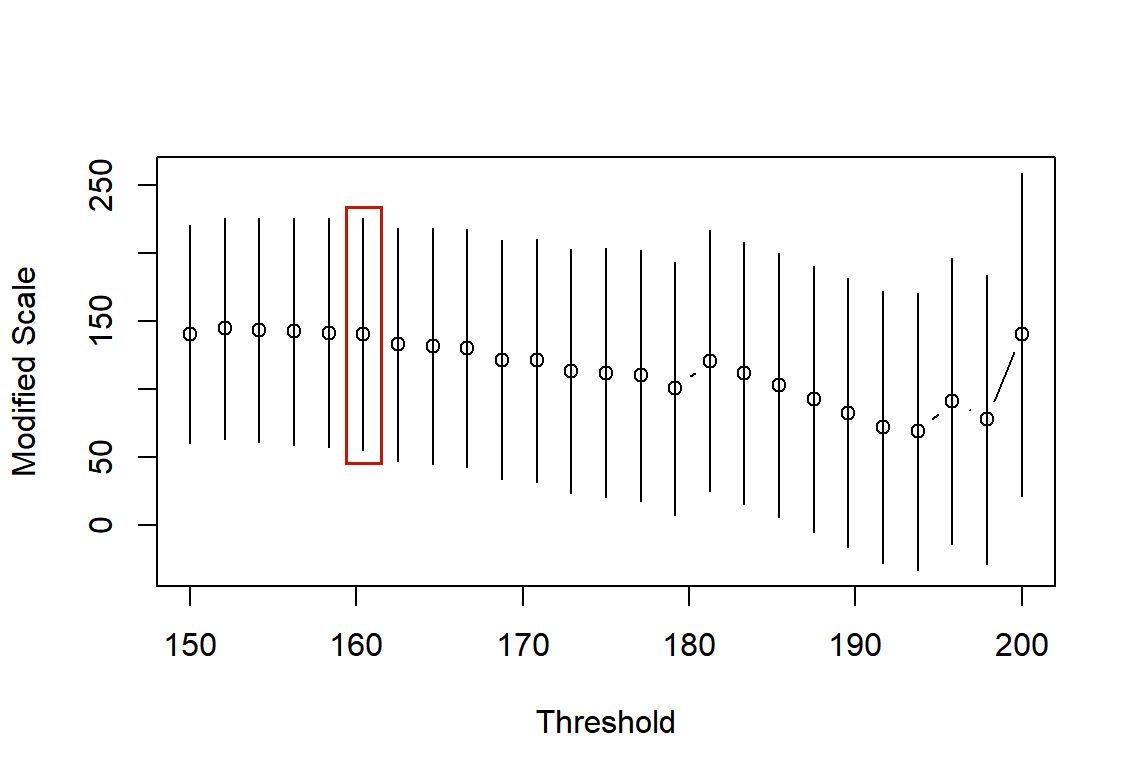}
  }
  \subfigure[]{\includegraphics[width=7cm,height=3.95cm]{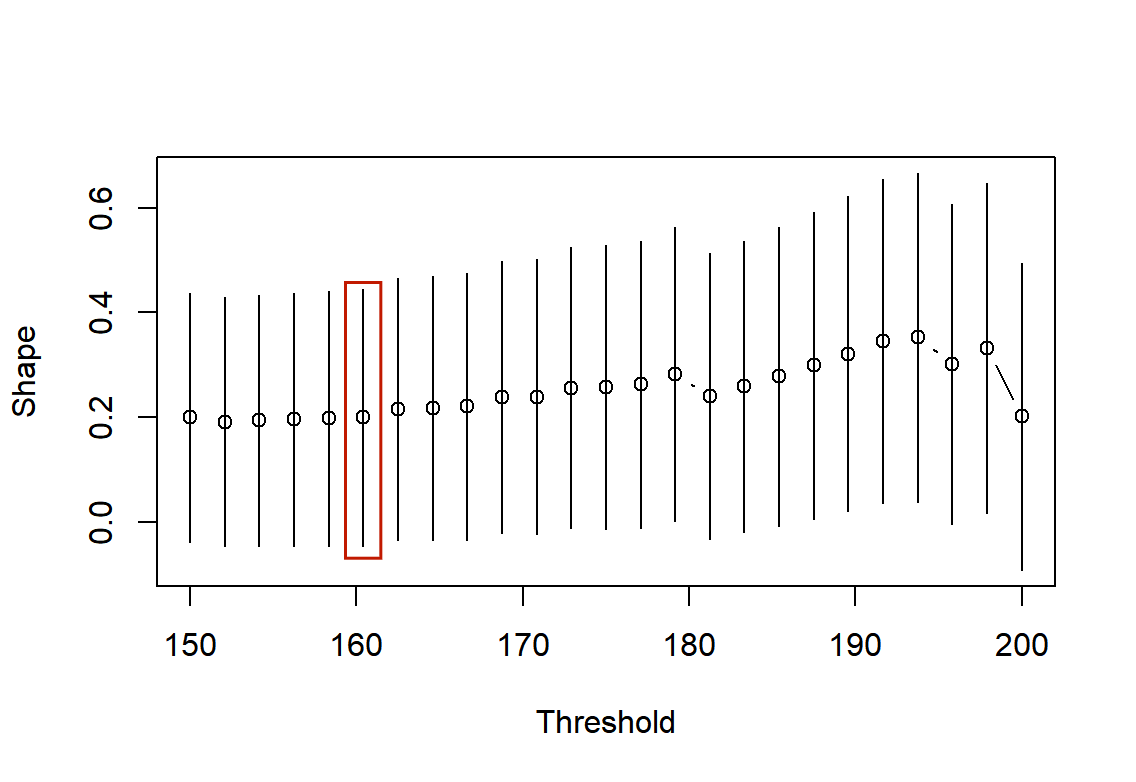}
  }
  
  \subfigure[]{\includegraphics[width=7cm,height=3.95cm]{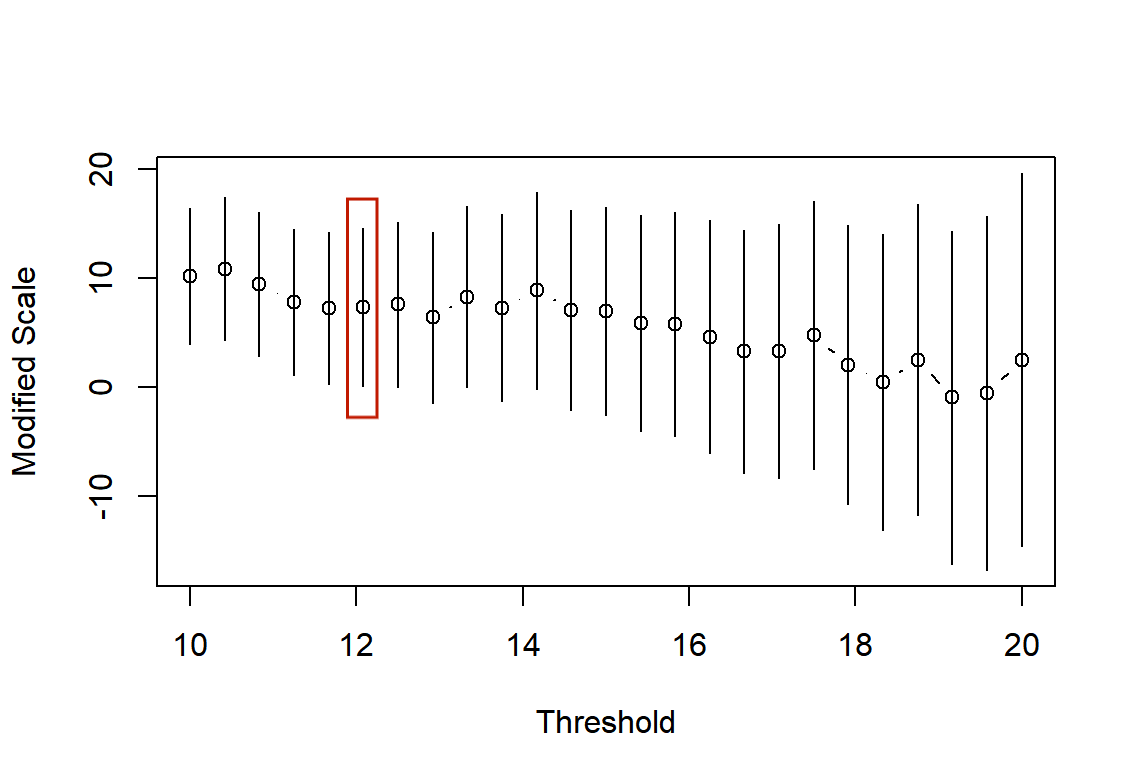}
  }
  \subfigure[]{\includegraphics[width=7cm,height=3.95cm]{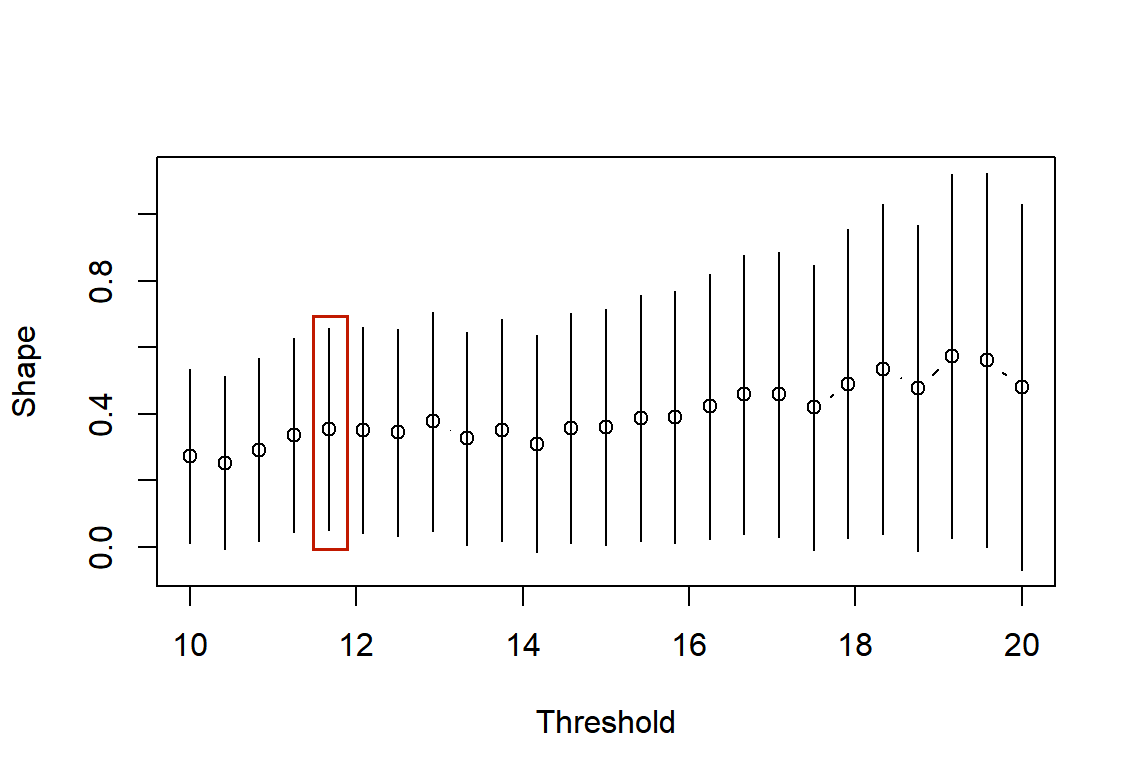}
  }
  
  \subfigure[]{\includegraphics[width=7cm,height=3.95cm]{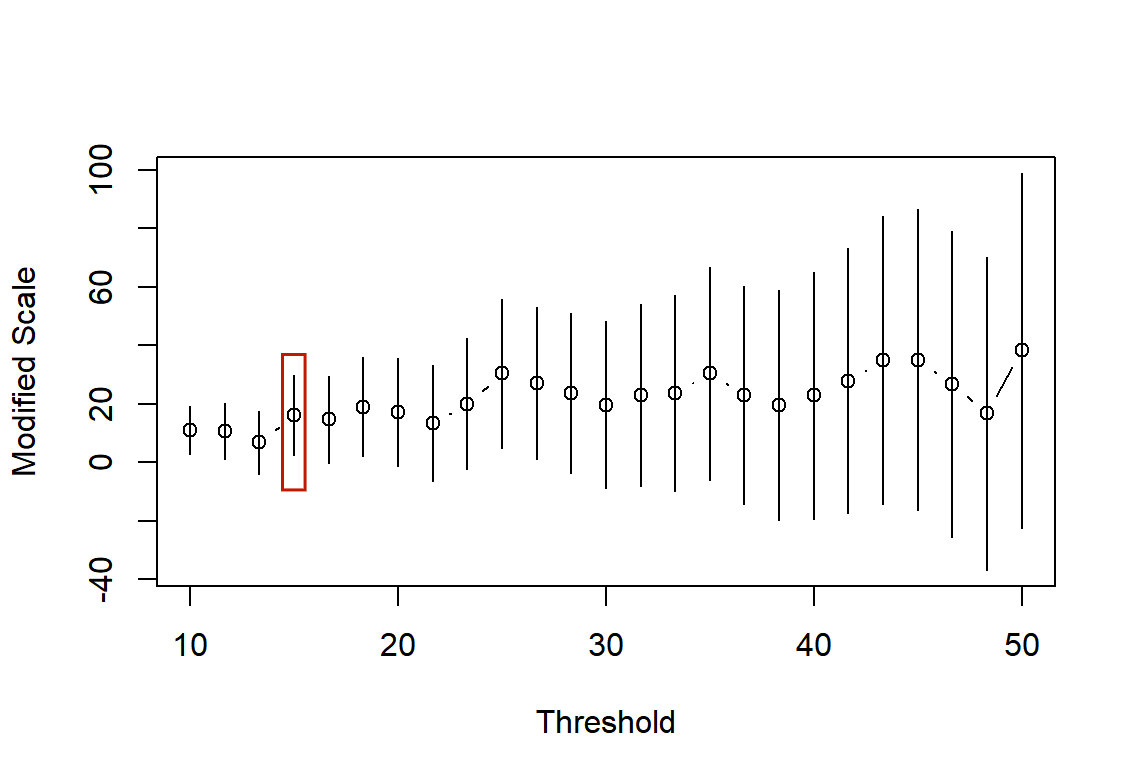}
  }
  \subfigure[]{\includegraphics[width=7cm,height=3.95cm]{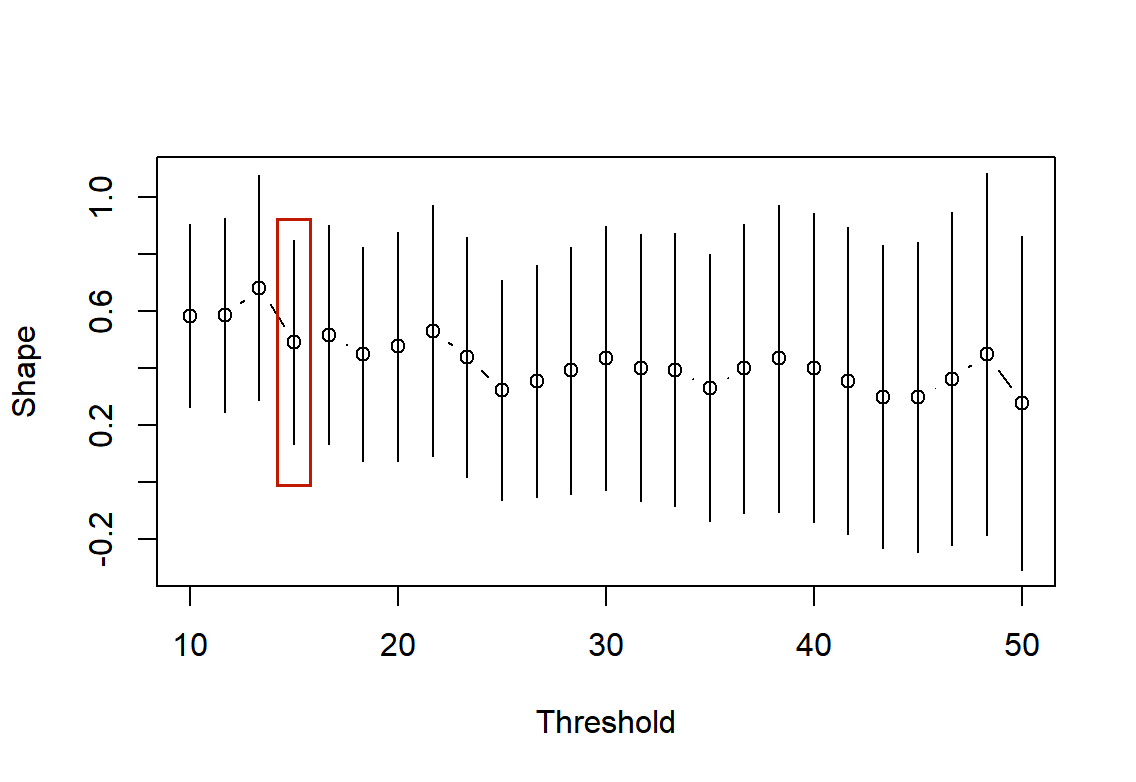}
  }
  \caption{Parameter stability plots on the top for affected population (AP), in the middle for crop affected area (CAA) and on the bottom for directed economic loss (DEL). The shorter the bars in (a, c, e) are, the more stable the estimations of the scale parameter $(\sigma)$ are, while (b, d, f) corresponds to the shape parameter $(\xi)$.}
  \label{Fig4}
\end{figure}

In order to determine the threshold level $(u_1, u_2, u_3)$ of  the triple of indicators $(X_1, X_2, X_3)$, we examine the mean residual life plots in Figure \ref{Fig3} (see Eq.\eqref{MEF} for details). We determine first the threshold $u_1$ for affected population $X_1$. The possible range of threshold is detected to be in the interval of $(150,200)$ according to the sample mean residual plot with a linear trend (see Figure \ref{Fig3}(a)), and then we investigate the stability of the estimations of the scale and shape parameters involved in the generalized Pareto model of the threshold-excesses of AP in Figure \ref{Fig4}(a)(b). Consequently, the attachment level $u_1 =160$ is determined for affected population (AP). Similar arguments give the attachment levels of $u_2 = 12$ for crop affected area (CAA) and $u_3 =15$ for directed economic loss (DEL).

\begin{figure}[htpb!]
  \centering
  \subfigure[]{\includegraphics[width=15cm,height=4cm]{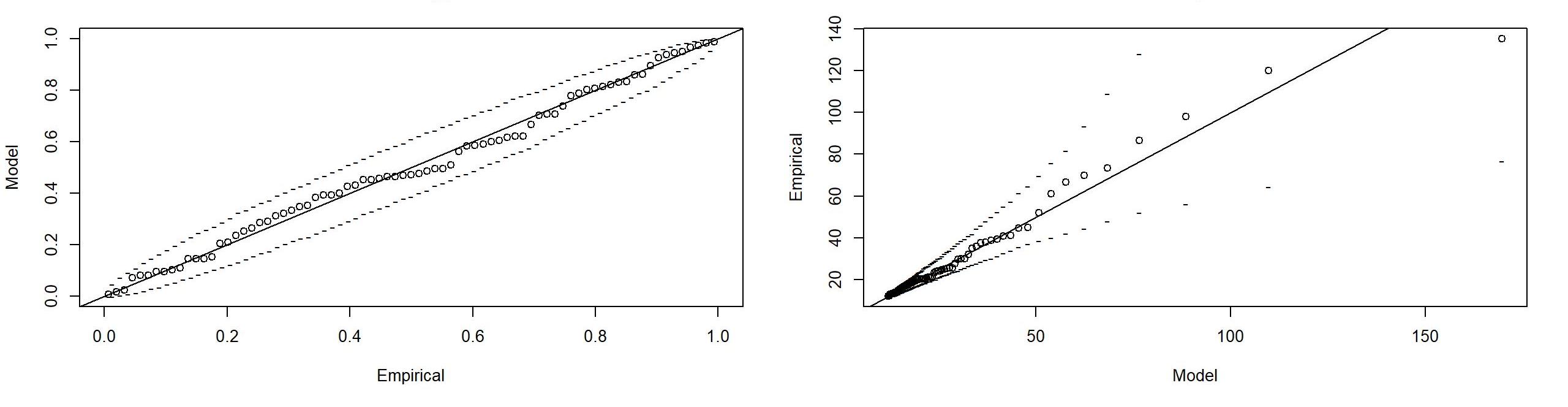}
  }
  \subfigure[]{\includegraphics[width=15cm,height=4cm]{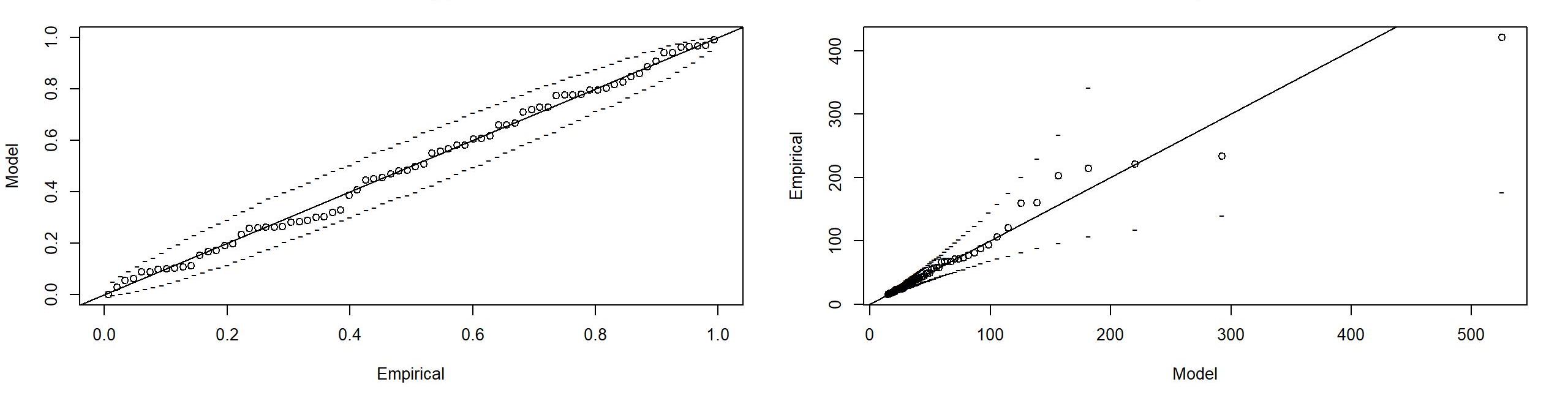}
  }
  \subfigure[]{\includegraphics[width=15cm,height=4cm]{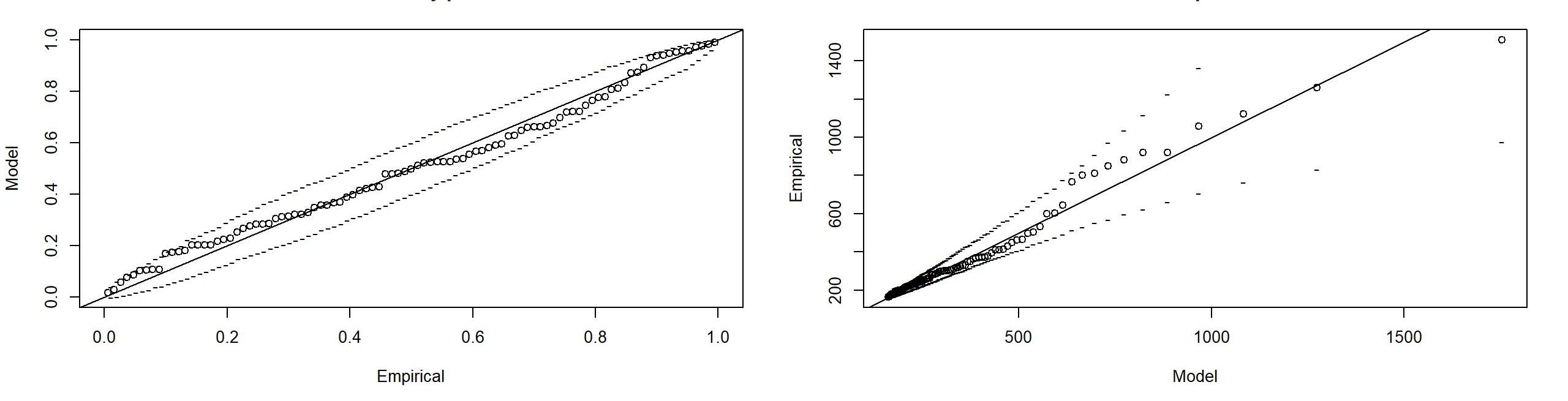}
  }
  \caption{The PP plots (left) and QQ plots (right) for (a) affected population (AP), (b) crop affected area (CAA) and (c) directed economic loss (DEL) subsequently.}
  \label{Fig5}
\end{figure}

We model the probability distribution of our trigger indicators using the full range of the data, with sufficient flexibility for separate control over bulk and tail features. Different from \cite{chao2018multiple}, we consider the Beta-GP models for the scaled non-exceedances and threshold-excesses, namely, 
\BQN
\label{Beta-GP-data}
\left\{
\begin{array}{ll}
X_{i} - u_i|X_{i}>u_i \sim G_{}(y) = 1-\left(1+\frac{\xi y}\sigma\right)^{-1/\xi},\, y>0, & \mbox{for threshold-excesses},\\
X_{i}^* = \frac{X_{i}- m_i}{u_i - m_i} \sim Beta_{\alpha, \beta}(y) = \frac1{B(\alpha, \beta)} y^{\alpha-1}(1-y)^{\beta-1},\, 0<y<1, & \mbox{for non-exceedances, i.e., $X_{i}<u_i$},  
\end{array}
\right. 
\EQN
where $(m_1, m_2, m_3) = (1.6, 0.005, 1.011)$ are the sample minima of the trigger indicators of AP, CAA and DEL given in Table \ref{Table1}, and $(u_1, u_2, u_3) = (160, 12, 15)$ are given above by the mean residual plots and parameter stability plots. All maximum likelihood estimations of the parameters involved in Beta-GP models are shown in Table \ref{Table4}. We see that both CAA and DEL possess heavy tails with 95\% confidence intervals of shape parameters as $(0.03, 0.64)$ and $(0.13, 0.84)$, while the estimated shape parameter for AP excesses is 0.197, showing a certain power decaying tail. Moreover, we examine the model goodness-of-fit using Chi-square test and all the $p$-values are larger than 0.246, confirming that the Beta-GP model agrees with the observed trigger indicators. Intuitively, Figure \ref{Fig5} tells us the appropriateness of the GP model of the threshold-exceedances for each trigger indicator since both PP plots and QQ plots show almost all points are around the straight line, indicating the sample \cl{quantile}
/empirical probability is rather close to the theoretical ones.

\begin{table}[ht]
\begin{center}
\caption{
Estimates of the scale $(\sigma)$ and shape $(\xi$) parameters in the $GP$ models of the threshold excesses and the two parameters $\alpha$ and $\beta$ in the Beta model $Beta_{\alpha, \beta}$ for the scaled bulk data below the threshold. The $p$-value based on Chi-square test confirms that the Beta-GP model fits the data well. }
\label{Table4}
\begin{tabular}{ccccccccc}
\toprule
& \multicolumn{5}{c}{GP}                           
& \multicolumn{3}{c}{Beta}                                \\ \cmidrule(r){2-6} \cmidrule(r){7-9}
    & Scale ($\sigma$) & 95\% CI & Shape ($\xi$) & 95\% CI  &$p$-value   & $\alpha$ & $\beta$    &$p$-value                          \\ \hline
 AP  & 173.369                   & (118.21, 226.38)          & 0.197                     & ($-$0.04, 0.44)   &  0.247      & 1.016 & 1.345 & 0.246  \\
 CAA & 11.771                    & (7.41, 16.13)             & 0.341                     & (0.03, 0.64)     & 0.283      &  1.076 &  1.687  & 0.299  \\
 DEL & 23.538                    & (13.96, 33.19)            & 0.492                     & (0.13, 0.84)   & 0.250          &  0.582 &  1.137 & 0.253  \\ 
\bottomrule
\end{tabular}
\end{center}
\end{table}

To conclude, the fitted distributions $F_i, i=1,2,3$ correspond to the trigger indicators of affected population (AP), the crop affected area (CAA), and the direct economic loss (DEL) is given below
\begin{equation}
\begin{normalsize}
F_i(x)= \begin{cases}1- \frac{n_{u_i}}{n} \overline G_{\xi_i, \sigma_i}(x-u_i), & x > u_i, \\ 
\frac{1}{B(\alpha_i, \beta_i)} \int_0^{(x-m_i)/(u_i-m_i)} t^{\alpha_i-1}(1-t)^{\beta_i-1} d t, & x\le u_i,\end{cases}
\end{normalsize}
\label{Beta-GP-model}
\end{equation}
where $n=245$, the threshold  $(u_1, u_2, u_3)$, the excess numbers $(n_{u_1}, n_{u_2}, n_{u_3})$, the sample minina $(m_1, m_2, m_3)$ and other parameters for Beta-GP models are given by Eq.\eqref{Beta-GP-data} and Table \ref{Table4}. 

Given the distributions of each trigger indicator in Eq.\eqref{Beta-GP-model}, it remains to discuss the dependence via nested Archimedean copula in the following section in order to illustrate our multiple-event triggered pricing mechanism in Section \ref{Section-3.5}.
\subsection{Dependence anaylsis of trigger indicators based on nested Archimedean copula}
\label{Section-3.3}
We shall investigate first the non-exchangeable dependence among the trigger indicators through Spearman rank correlation $\rho$. The Spearman $\rho$'s for (AP, CAA), (AP, DEL) and (CAA, DEL) are 0.771, 0.554 and 0.515, respectively.  We see that all the three pairs demonstrate certain degree of dependence, while a stronger dependence within (AP, CAA) than between them. This motivates us to describe the dependence using nested Archimedean copula, with an inner copula for (AP, CAA) with Archimedean parameter $\theta_1$, and DEL placed into the outer Archimedean copula with parameter $\theta_2$ (see Eq.\eqref{nestedArchimedean}).

Next, to model the dependence structure using nested Archimedean copula, we make a uniform distributed transformation of the raw data according to the marginal analysis in Section \ref{Section-3.2}, i.e., a straightforward application of Eq.\eqref{Beta-GP-model} gives 
$$\widetilde u_{ij} = F_i(x_{ij}), \quad i=1,2,3, \, j=1,2,\ldots, 245.$$
Finally, we examine the dependence structure using nested Archimedean copula with both inner and outer copula being one of the Gumbel, Clayton and Frank copula listed in Table \ref{Table1}, namely, we suppose that 
\BQN
\label{Frank}
(\widetilde u_{1j}, \widetilde u_{2j}, \widetilde u_{3j}) \stackrel{i.i.d.}{\sim} C_{outer}(C_{inner}(\widetilde u_{1j}, \widetilde u_{2j}; \theta_1), \widetilde u_{3j}; \theta_2),\quad j =1,\ldots, 245.
\EQN
We see from Table \ref{Table5} that the maximum likelihood estimates of inner parameter $\theta_1$ are all larger than that for the outer parameter $\theta_2$, agreeing with the stronger dependence within the inner variables than between them. Secondly, we will select the most suitable copula based on the Kolmognov-Smirnov (KS) test. We see that all nested Archimedean copula fits the data well and the best nested Gumbel copula is determined with the minimal KS test value of 0.04 and the maximum $p$-value of 0.61.

\begin{table}[htbp!]
\begin{center}
\caption{Maximum likelihood estimates of the parameters involved in the inner and outer copulas in Eq.\eqref{nestedArchimedean}. The Kolmognov-Smirnov (KS) test and its $p$-value indicates that the best model is the nested Gumbel copula.}
\label{Table5}
\begin{tabular}{ccccc}
\toprule
Inner and outer copula & {$\theta_1$ (inner) } & { $\theta_2$ (outer) } & {KS test } & {$p$-value} \\ \hline
{Gumbel} & { 44.68} & {22.80} & {0.06} & {0.12}\\
{Clayton} & {9.58} & {4.35} & {0.12} & {0.35}\\
{Frank} & {176.34} & {87.60} & \textbf{0.04} & \textbf{0.61}\\
\bottomrule
\end{tabular}
\end{center}
\end{table}

\subsection{Modelling of annual frequency of rainstorms in China}
\label{Section-3.4}
Note that our pricing mechanism is of discrete form. It follows from Eq.\eqref{Cashflow} that, the proportion of coupon and principal paid out depends not only on the severity of the disasters (thus the trigger indicators) but also on the independent annual frequency of disasters. Due to the changing climate, the occurrence of main rainstorms becomes more and more frequent.The purpose of this section is to model the intensity of annual main rainstorms during 1986-2020 in China  using auto-regressive moving average (ARMA) model so as to forecast the frequency in the next three years, i.e., 2021\cl{--}2023. The relevant data is from \href{http://cmdp.ncc-cma.net/} {\itshape National Climate Centre\upshape}.

\begin{figure}[ht]
  \centering
  \subfigure[]{\includegraphics[width=8cm,height=6cm]{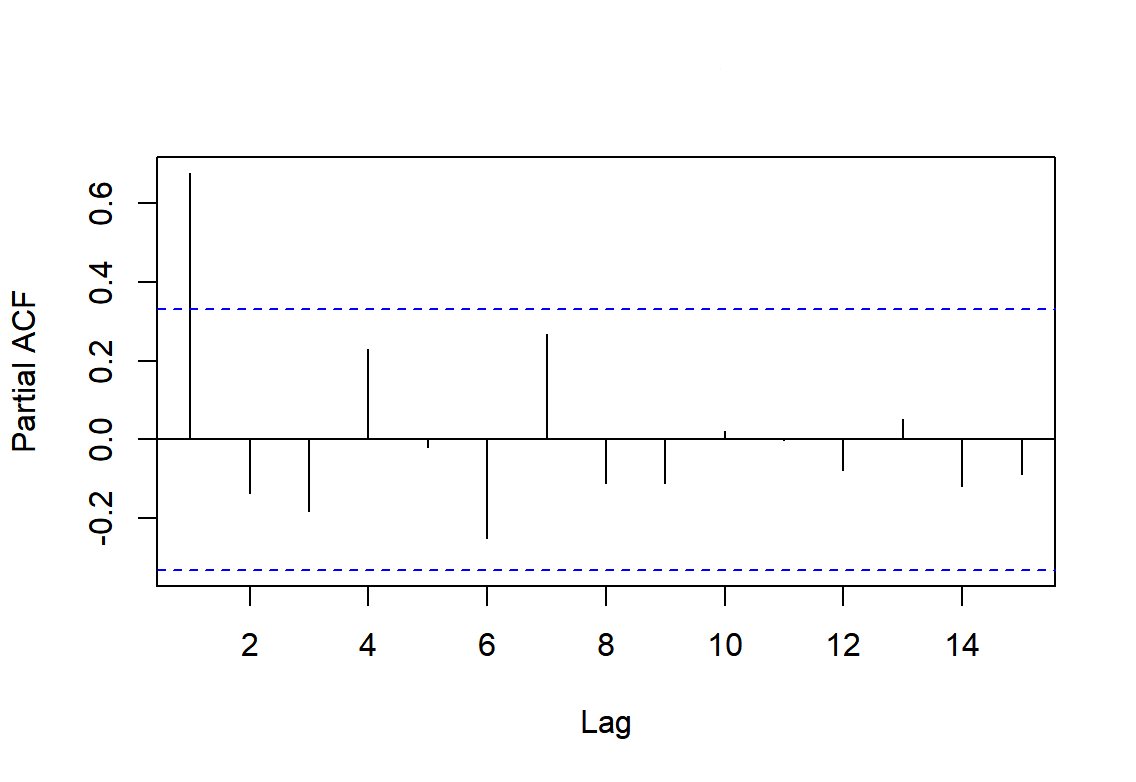}
   }
  \subfigure[]{\includegraphics[width=8cm,height=6cm]{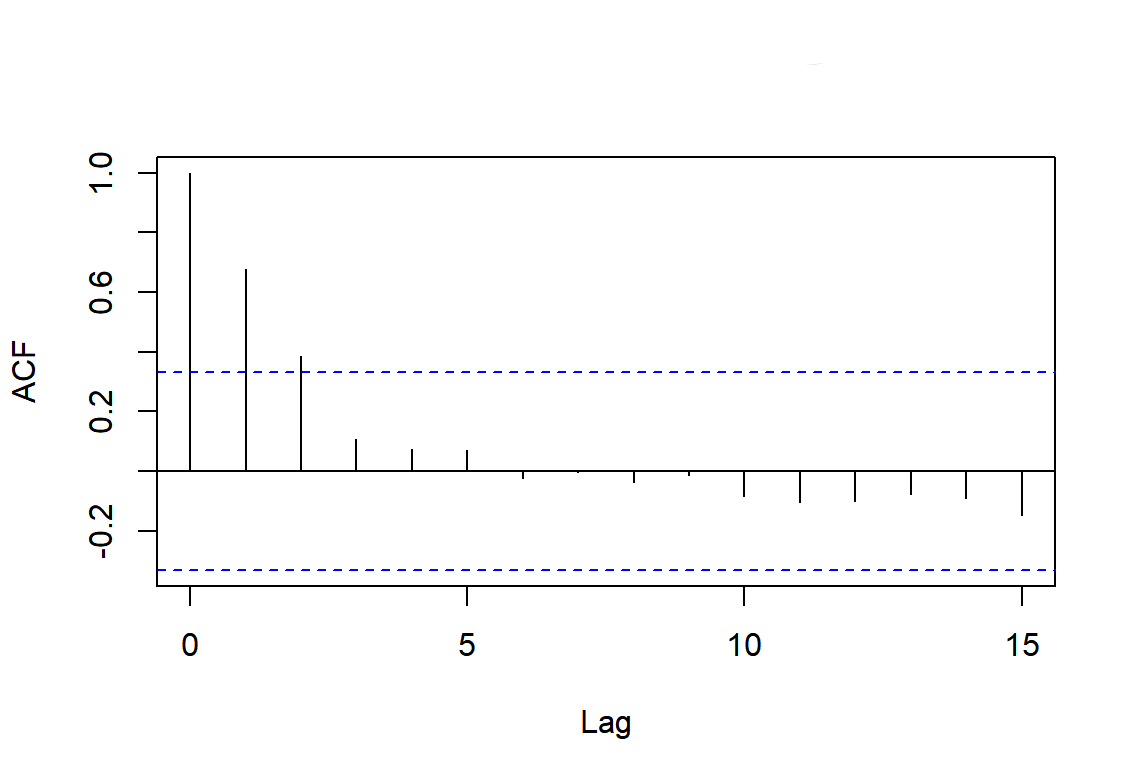}
    }
  \caption{(a) Partial autocorrelation function (PACF) and (b) autocorrelation function (ACF) diagram for annual intensity of rainstorms in China during 1985-2020.}
  \label{Fig6}
\end{figure}

The ARMA model for the intensity of annual rainstorms is as follows\cite{ibrahim2022multiple}:
$$ \Lambda_k=\mu +\phi_1  \Lambda_{k-1}+\ldots+\phi_p  \Lambda_{k-p}+\theta_1 e_{k-1}+\ldots+\theta_q e_{k-q}+e_k,$$
where $\Lambda_t=\Lambda(t+1)-\Lambda(t)$ represents the intensity measure of annual rainstorms in Year $j, j = 1, 2, \ldots, k$, and $p$ and $q$ respectively represent autoregressive, differentiation, moving-average order, and $e_k$ represents the random error. The assumptions in the ARMA$(p,q)$ are as follows:
\begin{itemize}
    \item
    The random errors
    $e_k$'s are independent and normal distributed with zero mean and constant variance, denoted by $e_k \stackrel{i.i.d.}\sim N(0, \sigma^2)$,
    \item
    The sequence $\Lambda_k$'s are
    weakly stationary, that is, $\forall k, \E{\Lambda_k}=\mu_\Lambda$, and $\operatorname{Var}\left(\Lambda_k\right)=\sigma_\Lambda^2$.
\end{itemize}
The stationarity assumption of $\Lambda_t $ is accepted according to the Augmented Dickey–Fuller (ADF) test with $p=0.05$. Next, we determine the partial autoregressive order $p=1$ and moving-average order $q=3$ respectively by  the partial autocorrelation function (PACF) and ACF plots in Figure \ref{Fig6}, since the PACF is cut-off at lag $p=1$ and the ACF at $q=3$. Thus, the ARMA(1,3) model is selected for modelling the annual intensity measure of main rainstorms in China and the led maximum-likelihood estimate of parameters involved are obtained as $(\phi_1, \theta_1,\theta_2,\theta_3)$ estimated as $(0.816, 0.268, 0.240,-0.748)$. 

\begin{figure}[ht]
  \centering
  {\includegraphics[width=10cm,height=6cm]{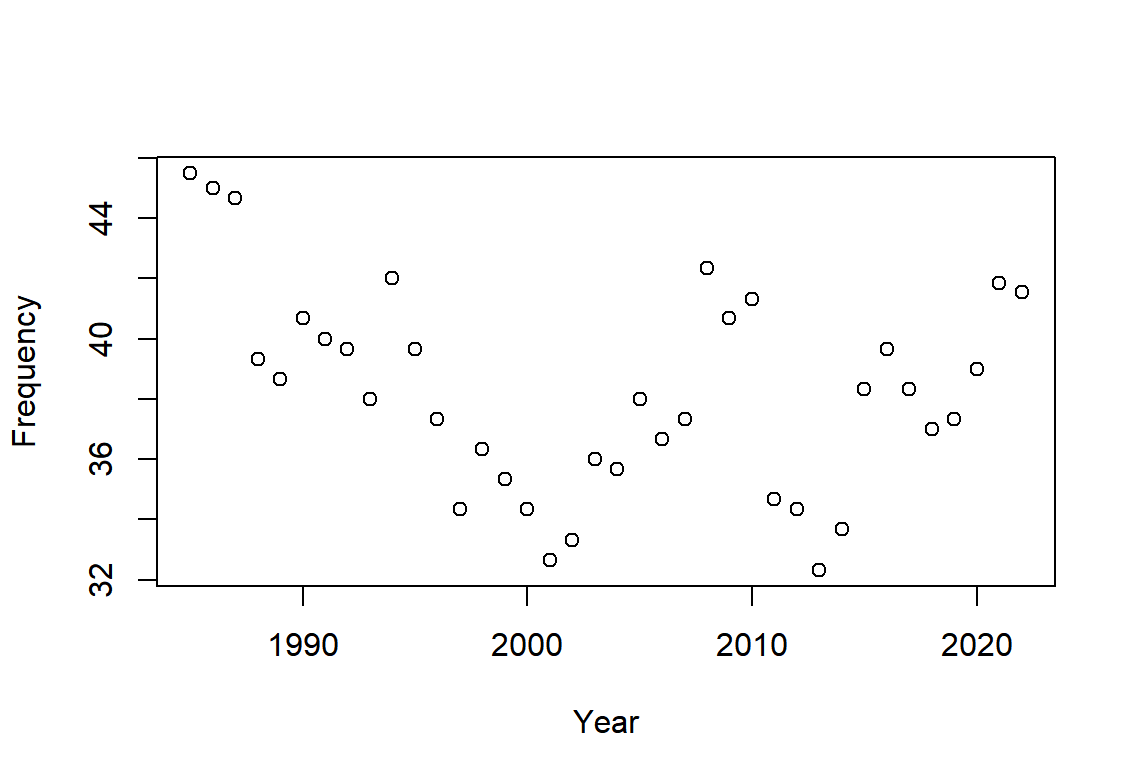}}
  \caption{Annual number of main rainstorms in China during 1985-2020, which data is from \href{http://cmdp.ncc-cma.net/} {\itshape National Climate Centre\upshape}.}
  \label{Fig7}
\end{figure}

The normality of $e_k$'s is accepted according to Jarque-Bera (JB) test with $p$-value of $0.08$, and its i.i.d. assumption is verified by the Ljung-Box (LB) test with $p=0.502$. Before we apply this model for prediction, the $F$-test confirms the goodness-of-fit with $p=0.003$. Consequently, the straightforward application of ARMA(1,3) to the intensity measures for 2021, 2022 and 2023 predicts $(41.86, 41.56, 39.39)$ a general upward trend in frequency over the next three years, as shown in Figure \ref{Fig7}, which will be used in Section \ref{Section-3.5} below for the simulation of pricing mechanism.

\subsection{Pricing of CAT bond}

\label{Section-3.5}
This section will focus on the CAT bond pricing based on our pricing mechanism in Eq.\eqref{PricingMechanism} and basic analyses of main rainstorms we conducted in Sections \ref{Section-3.1}-\ref{Section-3.4}. In what follows, we consider a $T$-year period CAT bond with a principal of $F=100$ yuan, fixed coupon rate $R=3.5\%$.
Note that \cl{the explicit expectation in} Eq.\eqref{PricingMechanism} is difficult to obtain. We thus estimate the price of CAT bond by using Monte Carlo simulation. The step of simulation is outlined as follows. 
\begin{itemize}
\item[(1)] Generate $T$ random numbers $N_1, N_2, \ldots, N_T$ independently from Poisson distribution with intensity measure $\Lambda_1 , \Lambda_2, \ldots, \Lambda_T$, standing for the number of main rainstorms in Year $1, 2, \ldots, T$. Here $(\Lambda_1,\Lambda_2, \ldots, \Lambda_T)$ can be provided by the ARMA(1,3) model in Section \ref{Section-3.4}. 
\item[(2)] Based on the marginal distribution in Eq.\eqref{Beta-GP-model} and the nested Frank copula in Eq.\eqref{Frank} with parameters given in Tables \ref{Table4} and \ref{Table5}, we generate $T$ sample of $(x_{1j}, x_{2j},x_{3j})$ of size $N_j$ for Year $j$ as 
$$x_{ij} = F_i^{-1}(\widetilde u_{ij}), \quad i=1, 2,3,\, j=\sum_{k=1}^{j-1} N_k+1, \ldots, \sum_{k=1}^{j} N_k.$$
\item[(3)] Here, we suppose that the stochastic interest rate process $\{r(t),\, t\ge0\}$ follows the CIR model stated in Eq.\eqref{CIR}, and the parameters involved are given by \cite{kladivko2007maximum} based on one-year seven-day interest rate data for 2021, available at the \href{https://www.shibor.org/}{\itshape Shanghai Interbank Offered Rate\upshape}. Initial interest rate is 6 January 2021, i.e.,   
\BQN\label{CIR-estimation}
\left\{
\begin{array}{l}
d r(t)=0.2\left(0.05-r(t)\right) d t+0.05 \sqrt{r(t)} d W_t,\\
r(0)=2.962\%.
\end{array}
\right.
\EQN
We simulate the interest risk to give the discount factor $p(t,T)$. 
\item[(4)] 
Given trigger level $(u_1, u_2, u_3)$, we calculate price of the CAT bond based on the cashflows in Eq.\eqref{Cashflow} and the interest risk model in Eq.\eqref{CIR-estimation}. 
\end{itemize}
We repeat all the simulation steps above {$m=10^4$} and get the sample mean of the price $P_t$ of a $T$-year CAT bond bought at year $t=0,1,\ldots, T-1$. In the following, we will discuss the pricing sensitivity in maturity period, trigger level and trigger indicators as well subsequently. 

\textbf{Price sensitivity in maturity period}. Table \ref{Table6} shows $T$-year CAT bond triggered by the triple of trigger indicators (AP, CAA, DEL) with trigger level $(u_1,u_2, u_3)$ being the sample 90\%-quantile. The bond price decreases in maturity period $T$ and further the downward trend in bond price gradually increases, as the bond is issued for a long maturity period, the future disaster might be more severe and frequent, causing larger magnitude of triggered events and thus a smaller proportion of coupon and principal retained. Consequently, a lower price is obtained for a medium-term CAT bond. Meanwhile, for a given maturity period $T=2$ or 3, the CAT bond might be \cl{sold} 
in different years during the bond issue period. We see that the price of a bond purchased in an earlier year is higher, as the potential risk of main rainstorms might be lower in the previous few years and the coupon and principal are more likely to remain. Therefore, the pricing mechanism is fairly attractive in terms of capital raising for the purpose of reinsurance.

\begin{table}[htbp!]
\caption{CAT bond pricing of main rainstorms in China purchased at Year $t$, maturating in Year $T=1,2,3$ based on Eq.\eqref{PricingMechanism}. Here the trigger level $(u_1,u_2, u_3)$ is taken as the sample 90\%-quantile of (AP, CAA, DEL).}
\label{Table6}
\centering
\begin{tabular}{ccccccc}
\toprule
& \multicolumn{1}{c}{$T=1$} 

& \multicolumn{2}{c}{$T=2$}     

& \multicolumn{3}{c}{$T=3$}                               \\ \cmidrule(r){2-2} \cmidrule(r){3-4} \cmidrule(r){5-7}
&  $t=0$ & $t=0$ & $t=1$ & $t=0$ & $t=1$ & $t=2$                           \\ \hline
 Bond price $P_t$ & 80.392 &  70.881 & 66.255 & 56.006 & 54.865 & 52.321 \\ 
\bottomrule
\end{tabular}
\end{table}

\begin{figure}[htbp!]
  \centering
  \subfigure[]{\includegraphics[width=7.3cm,height=4.5cm]{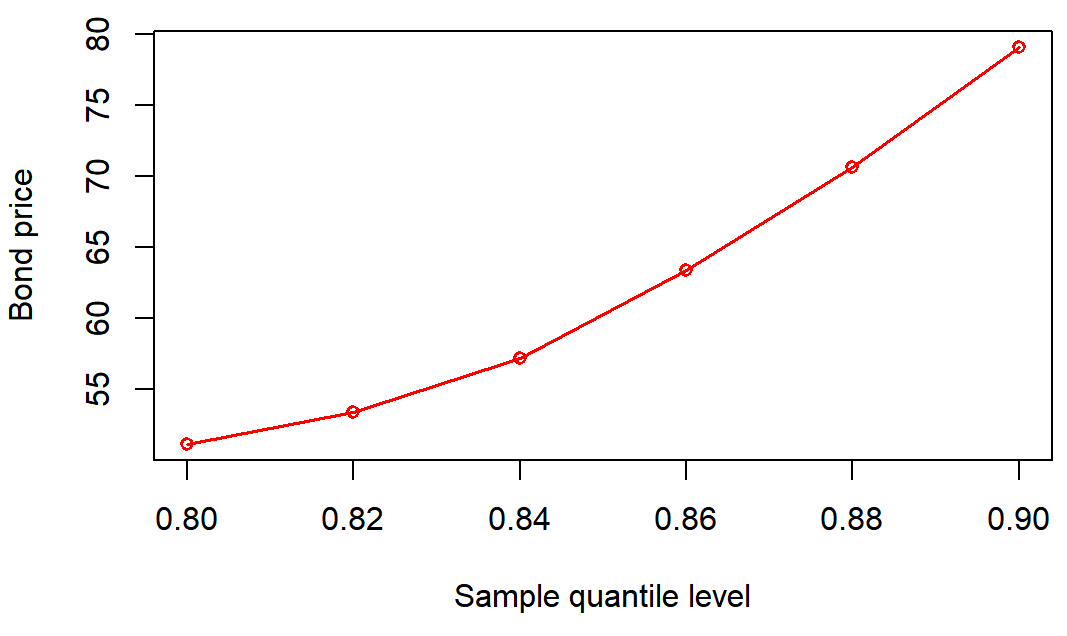}
 }
  \subfigure[]{\includegraphics[width=7.3cm,height=4.5cm]{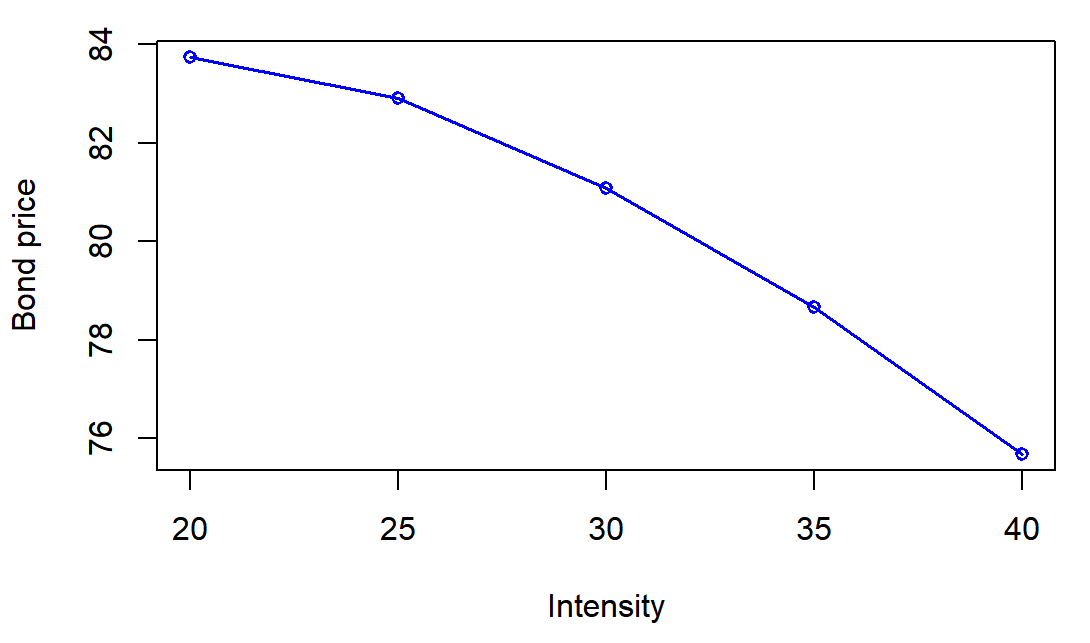}
  }
  \caption{Price sensitivity of one-year CAT bond in (a) trigger levels 
 (b) disaster intensity {measure $\Lambda_1=20,25,\ldots, 40$.} Here the trigger level in (a) is the sample quantile at level {$q=0.80, 0.82, \ldots,0.90$} for given intensity measure in Step 1, while in (b) we fix the trigger level as {its sample 90\%-quantile}.}
  \label{Fig8}
\end{figure}

\textbf{Price sensitivity in trigger level and intensity measure of main rainstorms.} 
An appropriate trigger level is of importance balancing the benefits between the investors and the bond issuers by means of the CAT bond price. In general, both the trigger levels and the intensity measures are determined by the severity and frequency of the potential disasters (here rainstorms), and it cause the change of the CAT bond price via the wiped-off coupon \& principal  in Eq.\eqref{Cashflow}.
Clearly, we see from Figure \ref{Fig8} that bond prices are positively correlated with trigger levels but negative associated with intensity measure. Indeed, 
as the trigger level increases, a bond is less likely triggered and thus the coupon and principal retention level will increase, leading to an increase in the CAT bond price. 
While if the main rainstorm occurs more frequently with larger intensity measure, the induced trigger indicators may accumulate a small amount of coupon \& principal  retained, causing thus  a decrease in bond prices. 

\textbf{Price sensitivity in the selection of trigger indicators}. As mentioned before, the multiple-event triggered CAT bond receives increasing attractions from both investors and CAT issuers. In Table \ref{Table7}, we compare the one-year bond prices with different pairs of trigger indicators under our pricing mechanism in Eq.\eqref{PricingMechanism}. Apparently, the three-event triggered CAT bond is issued with lower price than that with bivariate-event triggered ones. This is because the smaller trigger probability of concurrent trigger events leads to a small expectation of the discounted cashflows. Additionally, the principal might be half returned in case all three indicators are triggered (recalling the utilization of $\gamma_t$'s).

\begin{table}[htbp!]
\caption{Comparison of one-year CAT bond price with different selections of multiple-event triggers under our pricing mechanism in Eq.\eqref{PricingMechanism}.
}
\label{Table7}
\begin{center}
\begin{tabular}{ccccc}
\toprule
Trigger indicator &  AP-CAA-DEL &  AP-CAA & AP-DEL & CAA-DEL \\ \hline
Bond price        & 80.3922                            &   48.7558                                           &  48.7561                        &  48.7564\\ 
\bottomrule
\end{tabular}
\end{center}
\end{table}

\section{Conclusions and extensional discussions}

In this paper, a multiple-event triggered CAT bond pricing model is designed and a pricing formula is derived based on the copula--POT model. Our pricing model is more flexible and of more practical significance in terms of  possessing a dynamic association between coupon and principal paid off, with its functional magnitude changing in the potential disasters. This may provide certain reference for the pricing research and subsequent practical application of catastrophe bonds. Although the multi-event trigger mechanism involves more elements, and more complex processes and requires higher technical capabilities, the multiple-event triggered bonds receive increasing attraction due to its low moral hazard and trigger risk\cite{chao2018multiple,wei2022pricing,ibrahim2022multiple}. 
Meanwhile, it turns out that, the CAT bond price decreases in bond maturity period in the simulation,  and it is negatively correlated with trigger level and positively correlated with catastrophe intensity.

The pricing mechanism depends on the joint distribution of catastrophe loss indicators led by a common disaster. In real life, catastrophes involve multiple disasters over multiple regions. To build a network-based CAT bond pricing model taking spatio
-temporal extremes into account is a forthcoming research direction. Due to the unavailability of real trading data in the secondary market of CAT bonds, the constructed CAT bond pricing model only simulates bond prices through catastrophe loss data, without using real trading data to test its accuracy, which inspires another research direction in the near future.


\begin{thebibliography}{10}

\bibitem{aase1999equilibrium}
Knut Aase.
\newblock An equilibrium model of catastrophe insurance futures and spreads.
\newblock {\em The geneva papers on risk and insurance theory}, 24(1):69--96,
  1999.

\bibitem{anantapadmanabhan1971some}
CS~Anantapadmanabhan.
\newblock Some statistical aspects of catastrophic risks.
\newblock {\em ASTIN Bulletin: The Journal of the IAA}, 5(3):307--313, 1971.

\bibitem{chao2018multiple}
Wen Chao and Huiwen Zou.
\newblock Multiple-event catastrophe bond pricing based on {CIR}-{C}opula-{POT}
  model.
\newblock {\em Discrete Dynamics in Nature and Society}, 2018, 2018.

\bibitem{RePEc:eee:jmvana:v:100:y:2009:i:7:p:1521-1537}
Arthur Charpentier and Johan Segers.
\newblock {Tails of multivariate Archimedean copulas}.
\newblock {\em Journal of Multivariate Analysis}, 100(7):1521--1537, August
  2009.

\bibitem{chen2013pricing}
Junfei Chen, Guiyun Liu, Liu Yang, Quanxi Shao, and Huimin Wang.
\newblock Pricing and simulation for extreme flood catastrophe bonds.
\newblock {\em Water resources management}, 27(10):3713--3725, 2013.

\bibitem{cox1985intertemporal}
John~C Cox, Jonathan~E Ingersoll~Jr, and Stephen~A Ross.
\newblock An intertemporal general equilibrium model of asset prices.
\newblock {\em Econometrica: Journal of the Econometric Society}, pages
  363--384, 1985.

\bibitem{cox2000catastrophe}
Samuel~H Cox and Hal~W Pedersen.
\newblock Catastrophe risk bonds.
\newblock {\em North American Actuarial Journal}, 4(4):56--82, 2000.

\bibitem{cummins2008cat}
J~David Cummins.
\newblock Cat bonds and other risk-linked securities: state of the market and
  recent developments.
\newblock {\em Risk management and insurance review}, 11(1):23--47, 2008.

\bibitem{deng2020research}
Guoqu Deng, Shiqiang Liu, Li~Li, and Chushi Deng.
\newblock Research on the pricing of global drought catastrophe bonds.
\newblock {\em Mathematical Problems in Engineering}, 2020, 2020.

\bibitem{hofert2010sampling}
Marius Hofert.
\newblock {\em Sampling nested Archimedean copulas with applications to CDO
  pricing}.
\newblock PhD thesis, Universit{\"a}t Ulm, 2010.

\bibitem{ibrahim2022multiple}
Riza~Andrian Ibrahim, Herlina Napitupulu, et~al.
\newblock Multiple-trigger catastrophe bond pricing model and its simulation
  using numerical methods.
\newblock {\em Mathematics}, 10(9):1363, 2022.

\bibitem{karagiannis2016modelling}
N~Karagiannis, H~Assa, AA~Pantelous, and CG~Turvey.
\newblock Modelling and pricing of catastrophe risk bonds with a
  temperature-based agricultural application.
\newblock {\em Quantitative Finance}, 16(12):1949--1959, 2016.

\bibitem{kladivko2007maximum}
Kamil Klad{\'\i}vko.
\newblock Maximum likelihood estimation of the cox-ingersoll-ross process: the
  matlab implementation.
\newblock {\em Technical Computing Prague}, 7(8):1--8, 2007.

\bibitem{leppisaari2016modeling}
Matias Leppisaari.
\newblock Modeling catastrophic deaths using evt with a microsimulation
  approach to reinsurance pricing.
\newblock {\em Scandinavian Actuarial Journal}, 2016(2):113--145, 2016.

\bibitem{litzenberger1996assessing}
Robert~H Litzenberger, David~R Beaglehole, and Craig~E Reynolds.
\newblock Assessing catastrophe reinsurance-linked securities as a new asset
  class.
\newblock {\em Journal of Portfolio Management}, page~76, 1996.

\bibitem{ma2021return}
Ning Ma, Yanbing Bai, and Shengwang Meng.
\newblock Return period evaluation of the largest possible earthquake
  magnitudes in mainland {C}hina based on extreme value theory.
\newblock {\em Sensors}, 21(10):3519, 2021.

\bibitem{mcneil1997estimating}
Alexander~J McNeil.
\newblock Estimating the tails of loss severity distributions using extreme
  value theory.
\newblock {\em ASTIN Bulletin: The Journal of the IAA}, 27(1):117--137, 1997.

\bibitem{nowak2013pricing}
Piotr Nowak and Maciej Romaniuk.
\newblock Pricing and simulations of catastrophe bonds.
\newblock {\em Insurance: Mathematics and Economics}, 52(1):18--28, 2013.

\bibitem{pickands31}
J~Pickands.
\newblock Statistical inference using extreme order statistics.
\newblock {\em Ann. Statist}, 3:131, 1975.

\bibitem{reshetar2008pricing}
Ganna Reshetar.
\newblock Pricing of multiple-event coupon paying cat bond.
\newblock {\em Available at SSRN 1059021}, 2008.

\bibitem{sklar1973random}
Abe Sklar.
\newblock Random variables, joint distribution functions, and copulas.
\newblock {\em Kybernetika}, 9(6):449--460, 1973.

\bibitem{RN30}
Oldrich Vasicek.
\newblock An equilibrium characterization of the term structure.
\newblock {\em Journal of Financial Economics}, 5(2):177--188, 1977.

\bibitem{wei2022pricing}
Longfei Wei, Lu~Liu, and Jialong Hou.
\newblock Pricing hybrid-triggered catastrophe bonds based on copula-evt model.
\newblock {\em Quantitative Finance and Economics}, 6(2):223--243, 2022.

\bibitem{woo2004catastrophe}
Gordon Woo.
\newblock A catastrophe bond niche: Multiple event risk.
\newblock In {\em meeting of the NBER Insurance Project Group}, 2004.

\bibitem{zimbidis2007modeling}
Alexandros~A Zimbidis, Nickolaos~E Frangos, and Athanasios~A Pantelous.
\newblock Modeling earthquake risk via extreme value theory and pricing the
  respective catastrophe bonds.
\newblock {\em ASTIN Bulletin: The Journal of the IAA}, 37(1):163--183, 2007.

\end{thebibliography}
\end{document}